%% file: JOs_SIADS_revision.tex
\documentclass[final,onefignum,onetabnum]{siamonline220329} 

\usepackage{pstricks,paralist,datetime}

\input{shared.tex}
\input{./hudef.tex}
\input{./p2pdef.tex}

\allowdisplaybreaks

\begin{document}

\maketitle

\begin{abstract}
Spatiotemporal localized and extended structures associated with a subcritical finite wavenumber Hopf bifurcation are studied in the Purwins model (a three-variable FitzHugh-Nagumo version). Steady and time-dependent numerical continuation procedures are used to investigate snaking behavior of localized standing and traveling waves on the real line, and the results are corroborated using weakly nonlinear theory. The results shed light on the origin of so-called jumping oscillons and the organization of a nontypical homoclinic snaking structure of traveling pulses. The computations are extended to moderate size disks and used to identify wall-attached spots that travel along the disk boundary as well as wall-attached spots that oscillate in place and wall-attached jumping oscillons. The one-dimensional results are shown to be useful in interpreting the two-dimensional results. Domain-filling and mixed structures are also studied, demonstrating the variety of extended and localized states that emerge in two-space dimensions, ranging from periodic to disordered. The latter are potentially important for observations of waves in far-from-equilibrium media, such as those often observed in cell biology.
\end{abstract}



\section{Introduction}
Self-organization of spatiotemporal patterns in dissipative systems is broadly studied in both partial differential equations theory and in applications, ranging from biology through chemistry to ecology but including technology as well, for example, nonlinear optics, catalysis, batteries and photovoltaics. The emerging patterns can be spatially periodic or localized, either as fronts (or domain walls) between distinct patterns or take the form of the so-called dissipative solitons. The latter structures are spatially localized and may be time-independent equilibrium states or take the form of traveling (excitable) {\it pulses} (TPs), i.e., localized states that continuously propagate in space, or localized standing oscillations that are fixed in space but oscillate in time, i.e., \textit{oscillons}. Such localized oscillations were discovered a number of years ago in parametrically driven fluid and granular media, and classified into two types: standard oscillons, which oscillate about a steady background~\cite{umbanhowar1996localized} and reciprocal oscillons, which are embedded in a counter-oscillating background~\cite{blair2000patterns,yochelis2006reciprocal}.

\begin{figure}
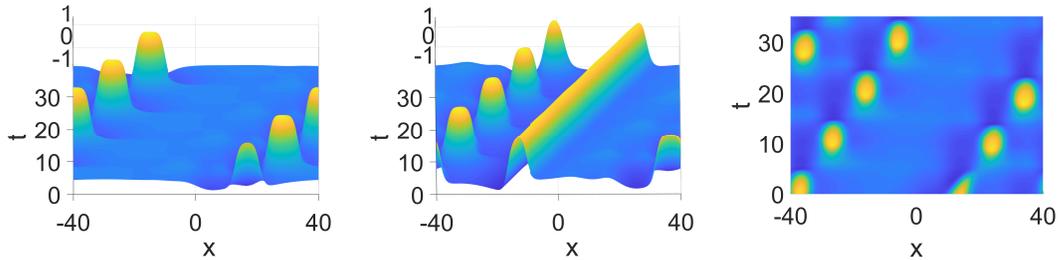

	\centering
	\ig[width=0.3\linewidth]{"fig1_1JO_3D-eps-converted-to"}
    \ig[width=0.3\linewidth]{"fig1_1JO1P_3D-eps-converted-to"}
    \ig[width=0.3\linewidth]{"fig1_2JO-eps-converted-to"}
	\caption{Space-time profiles of a jumping oscillon (1JO, left panel), a bound pair of a jumping oscillon and a traveling pulse (1JP-1TP, middle panel) and a bound pair of jumping oscillons (2JO, right panel). Reused from~\cite{knobloch2021origin}.}
	\label{fig:fig1}
\end{figure}

The so-called \textit{jumping oscillons} (JOs) are a third member of the ``oscillon family''. Unlike the two former types, {JOs were first observed in numerical simulations~\cite{YZE06} in one space dimension (1D) of the so-called Purwins system~\cite{schenk1997interacting} and subsequently in the form of jumping rings in a quasi-two-dimensional experiment on the BZ-AOT chemical system~\cite{cherkashin2008discontinuously}. Further examples were identified in mode-locked integrated external-cavity surface-emitting lasers~\cite{schelte2019third} and the associated first-principle model description~\cite{schelte2020dispersive}.} The term ``jumping'' refers to apparently discontinuous but periodic jumps in space, as shown in Fig.~\ref{fig:fig1} (left panel). While the mathematical principles behind standard and reciprocal oscillons have been studied extensively and are fairly well understood in 
1D~\cite{burke2008classification,alnahdi2014localized,mcquighan2014oscillons}, the origin of JOs in 1D has only recently been linked to strong temporal modulation of TPs~\cite{knobloch2021origin}; importantly, {the latter} study depended on developing a time-dependent numerical continuation method, details of which can also be found in Appendix~\ref{app:asec}. Subsequently, many additional states involving JOs were also discovered, including bound pairs of JO-TP [Fig.~\ref{fig:fig1} (middle panel)] and JO-JO [Fig.~\ref{fig:fig1} (right panel)]. These results have all been obtained in the Purwins system~\cite{schenk1997interacting}:
\begin{eqnarray}\label{eq:bm0}
    \nonumber  \pa_t u&=&k_1+k_2u-u^3-k_3v-k_4w+D_u\nabla^2 u,\\
    \theta \pa_t v&=&u-v+D_v\nabla^2 v,\\
    \nonumber  \vt\pa_t w&=&u-w+D_w\nabla^2 w, 
\end{eqnarray}
an extension of the canonical FitzHugh-Nagumo (FHN) activator-inhibitor model. Here $u$ is the activator and $v$, $w$ are two inhibitors acting on distinct time scales, $\theta$, $\vt$, respectively. The parameters $k_{1,2,3,4}$ represent rates while $D_{u,v,w}$ are diffusion coefficients, all assumed to be constant and positive.

In parallel, studies of spatially localized structures have been extended to two dimensions (2D)~\cite{avitabileSIADS2010,thieleNJP2019} and in particular to the behavior of such states on a disk~\cite{verschueren2021localized}, {illustrating} the important role played by the presence of a boundary: the linear stability problem describing instability of the homogeneous state has two types of eigenfunctions, those supported in the bulk and those supported by the {disk} boundary~\cite{goldsteinJFM1993}. The former yield bulk states while the latter yield wall-attached states. Both {eigenfunction types are periodic in the azimuthal direction and lead to structures that are periodic in this direction. In both cases secondary bifurcations may lead to spatially localized structures in the azimuthal direction. In particular, the {wall-attached state} may localize in the azimuthal direction, leading to a state localized in both radial and azimuthal directions, much as occurs in the 2D real Swift-Hohenberg equation \cite{verschueren2021localized}.} In practice, the bulk and wall modes are interlaced as a bifurcation parameter varies, leading to successive bifurcations to both mode types. Since these modes have very different spatial support their direction of branching may differ, so that, for example, the wall-modes may be subcritical while the bulk modes may be supercritical or vice versa. Of course, with increasing forcing, one expects the wall-attached states to invade the disk interior, and this is indeed what happens~\cite{verschueren2021localized}.

The work mentioned above was carried out using a gradient system, the Swift-Hohenberg equation, thereby excluding persistent dynamics. One motivation for the present work is to extend this type of study to the non-gradient case, specifically to the case where instability sets in as a symmetry-breaking Hopf bifurcation from a spatially homogeneous state $\bU_*\equiv (u_*,v_*,w_*)^{\rm T}$, that is, a Hopf bifurcation associated with a nonzero (azimuthal) wavenumber $m$. In the presence of periodic boundary conditions, whether on an interval in 1D or in the azimuthal angle $\Phi$ on a 2D disk, the resulting bifurcation problem is described by the theory for a Hopf bifurcation with $O(2)$ symmetry. In 1D this symmetry corresponds to spatial translations $x\mapsto x+\xi$ (modulo period) and reflections $x\mapsto -x$ about an arbitrary origin $x=0$; in the 2D case it corresponds to rotations $\Phi\mapsto \Phi+\xi$ (modulo period) and reflections $\Phi\mapsto -\Phi$. As a consequence, in 1D, the eigenvalues of $\bU_*$ are double, resulting in the simultaneous bifurcation to traveling (TWs) and standing waves (SWs, equal amplitude superpositions of left- and right- TWs)~\cite{knobloch1986oscillatory}. We call the analog of TW on a disk a rotating wave (RW). In either case, we refer to the corresponding bifurcation as a wave bifurcation. In two-species reaction-diffusion models this bifurcation is always preceded by an $m=0$ Hopf bifurcation and the role of subsequent wave bifurcations is therefore shielded{~\cite{ohta1996spontaneous,golubitsky2000target}}. This is no longer the case in systems of three or more reaction-diffusion equations, and we adopt here the system~\eqref{eq:bm0} as the model system of choice and study the dynamics of waves generated via a primary bifurcation of wave type. We are especially interested in exploring the relation between near-onset dynamics in 1D, described via coupled complex Ginzburg-Landau equations for the amplitudes of left- and right-traveling waves, and the near-onset wall-attached wave states in disks, and their interaction with bulk dynamics.
\begin{table}[tp!]
{\footnotesize
\caption{Solution types and acronyms.}\label{tab:simpletable}
\begin{center}
\begin{tabular}{|l|lc|}   \hline
\bf Acronym & \bf Meaning & \bf Geometry\\ 
\hline \hline
TWs & traveling waves&1D\\ 
TPs & traveling pulses & 1D \\ 
TPTs & traveling pulse trains & 1D\\ 
RSs & rotating spots & 2D\\ 
OSs & oscillating spots & 2D\\
RSTs & rotating spot trains & 2D\\ 
SWs & standing waves &1D \& 2D\\ 
LSWs &  localized standing waves & 1D \& 2D\\ 
JOs & jumping oscillons & 1D \& 2D\\
\hline      
HP&Hopf bifurcation onset &\\
PO&(time)-periodic orbit &\\
DNS&direct numerical simulation&\\
NBC&Neumann boundary condition&\\
PBC&periodic boundary condition&\\
BD&bifurcation diagram&\\
\hline
\end{tabular}
\end{center}
}
\end{table}

For definiteness, we restrict ourselves to the parameter regime already considered in~\cite{knobloch2021origin}, 
\begin{equation}\label{eq:opar}
    (k_2,k_3,k_4,\theta,\vt,D_u,D_v)=(2,10,2,50,0.5,2,25).
\end{equation}
We use $k_1$ as a control parameter, roughly in the range $k_1\in[-10,-6]$, while setting $D_w{=}100$ for 1D explorations (as in~\cite{knobloch2021origin}), but use $D_w{=}50$ for the 2D circular domain (i.e., on a disk), largely for numerical reasons. Table~\ref{tab:simpletable} summarizes the nomenclature used to refer to different solutions. We use SWs and LSWs as generic terms in 1D and 2D, but for 2D LSWs also use the more evocative term oscillating spots (OSs), in parallel with the use of rotating spots (ROs) in place of 1D TPs. Beyond the intriguing pattern-forming aspects of the present work, we hope that our results stimulate renewed interest in complex oscillatory dynamics often observed in large aspect ratio systems, including chemical systems~\cite{showalter2015chemical}, gas discharge media~\cite{purwins2010dissipative}, pipe flow~\cite{avila2023transition}, and even intracellular dynamics of interest in biology~\cite{horning2019three,kohyama2019cell,hoffmann2025corrections,kawamura2021mathematical,takada2026cell,ueda2026organized}.\\

\noindent
The paper is organized as follows: 
{\begin{description}
\item [In Section~\ref{sec:1D},]
we overview and extend the results from~\cite{knobloch2021origin}, focusing on the homoclinic snaking properties of standard oscillons and traveling pulses in 1D, including a weakly nonlinear analysis to uncover the origin of standard oscillons. The latter were not discussed in~\cite{knobloch2021origin} owing to the focus on jumping oscillons and constraints imposed by numerical continuation in both time and space which required relatively small domains. We also study a set of model amplitude equations, the coupled complex Ginzburg-Landau equations (CCGLEs), with coefficients derived from \eqref{eq:bm0}, to compute small amplitude spatially localized standing and traveling structures and unveil the role played by the group speed and the imaginary parts of the CCGLE coefficients.
\item [In Section~\ref{sec:2D},]
we focus on both spatially localized and spatially periodic solutions on disks. We rely on numerical continuation and direct numerical simulations (DNS) to study the similarities and differences between 1D and 2D patterns on disks with radii $R=20$ and $30$. Owing to the spatio-temporal complexity of many oscillating patterns, we report our results in the form of space-time plots along either the perimeter $\xi\in [-\pi R,\pi R]$ or the diameter $x\in [-R,R]$ (at $y=0$), depending on the state, together with selected snapshots and supplementary movies. \textit{We strongly encourage the reader to consult these movies.}
\item [In Section~\ref{sec:conclusions},]
we conclude the paper with a discussion and a brief outlook for the broader significance of our results for applications in which complex time-dependent solutions are frequently observed.
\item [{\rm Three} Appendices] provide technical details of derivations and the numerical implementation. In Appendix~\ref{app:asec} we comment on the numerical continuation technique we use, in Appendix~\ref{app:amp_eq} we give details of the derivation of the CCGL equations, while Appendix~\ref{app:disp_rel} shows how onset parameter values and amplitude equation coefficients can be computed semi-analytically from the dispersion relation.
\end{description}}
\begin{figure}[tp!]
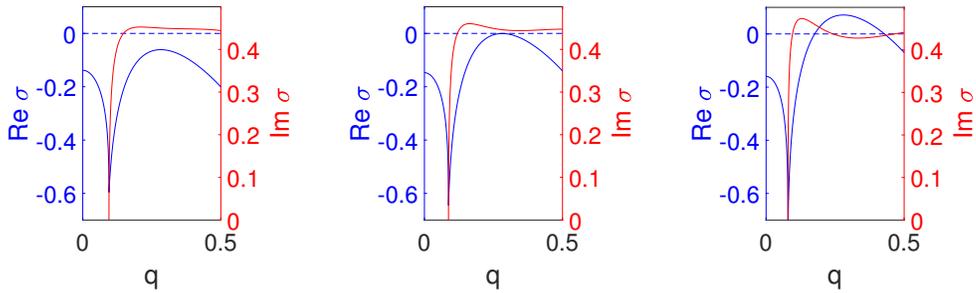

	\centering
	\ig[width=0.25\linewidth]{"disp_qc_below2-eps-converted-to"}\hspace{0.2in}
	\ig[width=0.25\linewidth]{"disp_qc2-eps-converted-to"}\hspace{0.2in}
	\ig[width=0.25\linewidth]{"disp_qc_above2-eps-converted-to"}
	\caption{Dispersion relations $\sigma(q)$ summarizing the linear stability properties of $\bU_*$ for various values of $k_1$, from left to right: $k_1=-7.9<k_{1c}$ (stable), $k_1=-7.585\simeq k_{1c}$ (critical), $k_1=-7.2>k_{1c}$ (unstable). The real (imaginary) parts of $\sigma$ are indicated in blue (red) on the left (right) axes. Note that the decaying ${\rm Re}\,\sigma$ from $q=0$ is in fact accompanied by ${\rm Im}\, \sigma=0$.}
	\label{fig:fig2}
\end{figure}

\section{Spatially periodic and localized solutions in 1D}\label{sec:1D}
We start with a brief review of the finite wavenumber Hopf (or wave) instability of the uniform solutions of the Purwins system~\eqref{eq:bm0}. For the parameters in~\eqref{eq:opar}, the Purwins system has a uniform steady state~\cite{YZE06,knobloch2021origin}
$\bU_* = (u_*,v_*,w_*)^{\rm T} = (u_*,u_*,u_*)^{\rm T}$, where 
\[
	u_*=\sqrt[3]{\Delta + k_1/2} - \sqrt[3]{\Delta - k_1/2}, \quad  \text{and} \quad \Delta=\sqrt{k^2_1/4+\bra{k_3+k_4-k_2}^3/27}. 
\]
Linear stability of $\bU_*$ in 1D is determined through the dispersion relation $\sigma(q)$ obtained on inserting 
\[
\bU-\bU_*\propto e^{\sigma t\pm iqx},
\] 
into \reff{eq:bm0} linearized about $\bU_*$. Here $\Re \, \sigma$ and $\omega \equiv\Im \, \sigma$ are the growth rate and frequency of the perturbation and both depend on its wavenumber $q$. Thus $\bU_*$ is linearly stable if $\Re \, \sigma(q) < 0,\, \forall q$, as demonstrated in Fig.~\ref{fig:fig2}(a). The instability onset is found at $k_{1c}\simeq -7.59$ at which $\Re \, \sigma(q_c)=0$ and $\text{d} \,[\Re \,\sigma]/\text{d}q=0$ at a critical wavenumber $q=q_c\simeq 0.28$, see Fig.~\ref{fig:fig2}(b). Hence $\bU_*$ is linearly unstable for $k_1>k_{1c}$, as shown in Fig.~\ref{fig:fig2}(c). The instability at $k_1=k_{1c}$ is of finite wavenumber Hopf or wave type (since $\omega\equiv\Im\, \sigma \neq 0$) and, as already mentioned, gives rise simultaneously to two time-periodic states, traveling and standing waves~\cite{knobloch1986oscillatory}. Note that the group speed at onset is small but not negligible, $|s_g(k_{1c})|\equiv |{\rm d}\omega/{\rm d}q|_{k_{1c}}| \simeq 0.074$. 

To determine the bifurcating TW and SW solutions and their criticality, we employ a multiple scale expansion together with the technique of reconstitution {(see Appendix~\ref{app:amp_eq} for details, including the method of reconstitution~\cite{spiegel1981physics,roberts1985introduction} required for the derivation)}. We start by expanding $\bU\equiv (u, v, w)^{\rm T}$ in the small parameter $\delta^2\propto |k_{1c}-k_1|\ll1$ designating the distance from the instability onset:
\begin{equation}\label{eq:perturb}
  \bU = \bU_*(\delta) + \delta \bU_1(x,X_i,t,T_i) + \delta^2 \bU_2(x,X_i,t,T_i) + \delta^3 \bU_3(x,X_i,t,T_i) + \mathcal{O}\bra{\delta^4}, 
\end{equation}
where $\bU_*$ represents the base state and $X_i=\delta^i x$, $T_i=\delta^i t$, $i=1,2$, are slow spatial and temporal scales (with respect to $x$ and $t$) corresponding to large scale, low frequency modulation of the bifurcating state {according to}: 
\begin{align*}\label{eq:slowamp}
    u_1 &= A_{\text{L}1}(X_1,X_2,T_1,T_2)e^{i(\omega_ct + q_cx)} + A_{\text{R}1}(X_1,X_2,T_1,T_2)e^{i(\omega_ct - q_cx)} + c.c.,\\
	v_1 &= A_{\text{L}2}(X_1,X_2,T_1,T_2)e^{i(\omega_ct + q_cx)} + A_{\text{R}2}(X_1,X_2,T_1,T_2)e^{i(\omega_ct - q_cx)} + c.c.,\\
	w_1 &= A_{\text{L}3}(X_1,X_2,T_1,T_2)e^{i(\omega_ct + q_cx)} + A_{\text{R}3}(X_1,X_2,T_1,T_2)e^{i(\omega_ct - q_cx)} + c.c.,
\end{align*}
where c.c. stands for a complex conjugate, $q_c$ and $\omega_c$ are the wavenumber and frequency at the instability onset, together with the eigenrelations
\[
A_{{\rm R,L};2}= \frac{1}{1 + D_v q_c^2 + i\omega_c\theta}A_{{\rm R,L};1}\equiv a_2A_{{\rm R,L};1},\quad
A_{{\rm R,L};3}= \frac{1}{1 + D_w q_c^2 + i\omega_c\vt}A_{{\rm R,L};1}\equiv a_3A_{{\rm R,L};1}.	
\]
When the group speed $s_g$ is finite but small the system is formally described by a pair of coupled amplitude equations, appropriately rescaled, for the amplitudes $A_{\text{L1,R1}}(X_1,X_2,T_1,T_2) \to A_{\text{L,R}}(X\equiv X_1+\delta X_2,T \equiv \delta^{-1}T_1+T_2)$ whose form follows from symmetry considerations~\cite[\S7.6]{hoyle2006pattern}:
\begin{subequations}\label{eq:final_ampltd}
	\begin{align}
\partial_T A_{\text{L}} - {S}_g \partial_X A_{\text{L}} &= \lambda\alpha A_{\text{L}} + \bra{1 + i\beta} \partial_X^2 A_{\text{L}} + \Bra {\bra{1 + i\gamma} |A_{\text{L}}|^2 + \eta |A_{\text{R}}|^2 } A_{\text{L}}, \\[1.5ex]
\partial_T A_{\text{R}} + {S}_g \partial_X A_{\text{R}} &= \lambda\alpha A_{\text{R}} + \bra{1 + i\beta} \partial_X^2 A_{\text{R}} + \Bra {\bra{1 + i\gamma} |A_{\text{R}}|^2 + \eta |A_{\text{L}}|^2 } A_{\text{R}}, 
	\end{align}
\end{subequations}
where $S_g = \delta^{-1}\kappa s_g \sim \mathcal{O}(1)$ is the normalized group speed. The coefficients in these equations (the CCGLE) are computed in Appendix~\ref{app:amp_eq}, leading to $\kappa=1/\sqrt{(1/2)|{\rm d}^2 {\rm Re}\, \sigma(k_{1c})/{\rm d}q^2|}\simeq 0.472$, while the remaining coefficients are $\alpha\equiv\alpha_r+i\alpha_i\simeq 0.189-0.025 i$, $\beta\simeq-0.1616,\gamma\simeq-5.6197$ and $\eta\equiv\eta_r+i\eta_i\simeq11.3644 - 1.0806 i$, cf.~\cite{crossPRL1986} and references therein. We set the remaining parameter to be $\lambda=-1$, corresponding to the subcritical regime $k_1<k_{1c}$, a regime known to support spatially localized structures. This procedure leaves the parameter $\delta$ in the group speed term only, implying that changes in the distance $|k_{1c}-k_1|$ from threshold are equivalent to changes in the normalized group speed $S_g$. In the following, we therefore investigate the properties of \eqref{eq:final_ampltd} as a function of $S_g$, for different values of the remaining ($\delta$-independent) coefficients. {Equations~\eqref{eq:final_ampltd} can also be derived using center manifold methods~\cite{roberts2015macroscale,roberts2017slowly,bunder2021nonlinear}.}

Equations \eqref{eq:final_ampltd} are not asymptotic (the limit $\delta\to0$ can only be taken when $s_g$ is itself $\mathcal{O}(\delta)$) and so must be viewed as {\it model equations} that are valid when $\delta\sim \mathcal{O}(s_g)$ and so apply outside the asymptotic regime $\delta\to0$; here, the relation 
\begin{equation}\label{eq:scaling}
\delta S_g=\kappa s_g=\frac{s_g}{\sqrt{(1/2)|{\rm d}^2 {\rm Re}\, \sigma(k_{1c})/{\rm d}q^2|}} \simeq 0.035    
\end{equation}
relates assumed values of $|S_g|$ to the corresponding values of $\delta$, i.e. the distance from threshold. We mention that the generic case, $s_g \sim \mathcal{O}(1)$, leads to a two time scale problem, with advection on a $\mathcal{O}(\delta^{-1})$ time scale and diffusion on the longer $\mathcal{O}(\delta^{-2})$ time scale. This case is described by distinct asymptotic equations whose form depends on the distance from onset: if $|k_{1c} -k_1|\sim \mathcal{O}(\delta^2)$ the leading order description of the evolution of $\mathcal{O}(\delta)$ amplitudes is a pair of linear hyperbolic equations on the advective time scale, leading at next order to a pair of coupled parabolic nonlinear but nonlocal equations~\cite{KnoblochDeLucaNON3,schneider1997justification}. On the other hand, if $|k_{1c} -k_1|\sim \mathcal{O}(\delta)$ the amplitudes are necessarily larger, $\mathcal{O}(\sqrt{\delta})$, and the dynamics are then described at leading order in the advective time scale by a nonlinear but local hyperbolic system of equations with no diffusion~\cite{MartelVegaNON1996,MartelVegaNON1998}. Both of these asymptotic descriptions differ from the model equations \eqref{eq:final_ampltd} but are not pursued in this work. In the following, we show that despite its nonasymptotic nature, the resulting model is nonetheless quantitatively useful.
\begin{figure}[tp!]
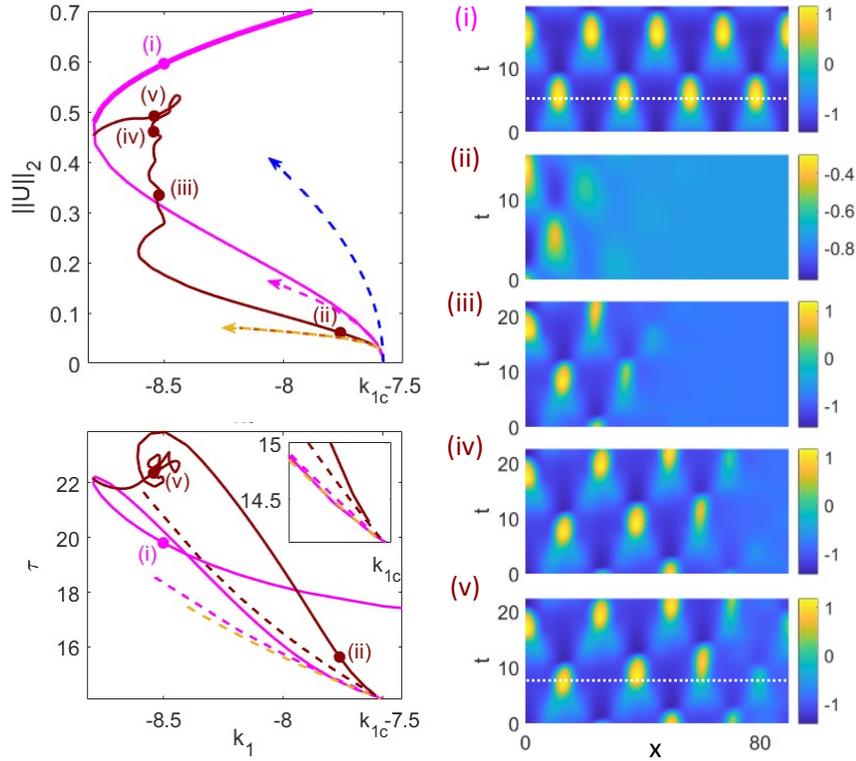

	\centering
	\ig[width=0.75\linewidth]{"fig3-eps-converted-to"}
        \caption{Bifurcation diagram (left panels) obtained via continuation with NBCs on a domain of length $L{=}4\cdot 2\pi/q_c\simeq 89.76$, showing branches of uniform standing waves (SWs, magenta) and spatially localized standing waves (LSWs, brown) in terms of the norm $\|U\|_2$ and the temporal period $\tau$, together with selected space-time plots at locations (i)-(v). Here, $x=0$ indicates the center of the LSW; the solutions must be reflected in $x{=}0$ to obtain the full LSW solution. The dotted white lines in panels (i) and (v) show that the SW oscillate in phase at every location while this is no longer the case for the LSW where the outer regions lag behind the center. The corresponding results from the amplitude equations are shown in dashed lines: magenta line is from~\eqref{eq:SW_ampltds}, while the orange line is obtained from~\eqref{eq:clSW} (i.e., with $S_g=0$) and the brown line is from continuation of~\eqref{eq:final_ampltd}; in the top left panel the orange and the brown lines lie one on top of one another. For reference, we also plot the TW branch from~\eqref{eq:TW_ampltds} [dashed blue line, see also Fig.~\ref{fig:appTW}]. Thick (thin) lines indicate stable (unstable) states.
        }
	\label{fig:fig3}
\end{figure}

{Equations~\eqref{eq:final_ampltd}} have solutions in the form of spatially uniform $(A_{TW},0)$ or $(0,A_{TW})$ {corresponding to TWs in~\eqref{eq:bm0},} 
\begin{equation}\label{eq:TW_ampltds}
A_{TW} = \sqrt{\alpha_r} e^{i \Omega_{TW} T},\qquad \Omega_{TW} = -\alpha_i+\gamma\alpha_r,
\end{equation}
and spatially uniform $(A_{SW},A_{SW})$ {corresponding to SWs in~\eqref{eq:bm0},} 
 \begin{equation}\label{eq:SW_ampltds}
A_{SW} = \sqrt{\frac{\alpha_r}{1 + \eta_r}}e^{i \Omega_{SW} T},\qquad \Omega_{SW} = -\alpha_i+\frac{\gamma+\eta_i}{1+\eta_r}\alpha_r,
\end{equation}
 both of which bifurcate subcritically from $\bU_*$ (Fig.~\ref{fig:fig3}, left panels, dashed lines), in agreement with numerical continuation of~\eqref{eq:bm0} (solid lines); a quantitative comparison between the approximate TW solution and the numerical TW solution is provided in Appendix~\ref{app:amp_eq}. It follows that the temporal period $\tau$ (for SW) and the speed $s$ (for TW) are given by
\begin{equation}
\tau = \frac{ 2\pi}{\omega_c + \delta^2 \Omega} ,\qquad s = \frac{\omega_c + \delta^2 \Omega_{TW}}{q_c},
\end{equation}
where $\Omega$ represents $\Omega_{SW}$ for uniform standing waves or $\Omega_{LSW}$ for localized standing waves (LSWs), see \eqref{eq:clSW}. These approximations are included in Figs.~\ref{fig:fig3} and~\ref{fig:appTW} as dashed lines.

We now turn to the organization of the SWs and LSWs further from onset (Fig.~\ref{fig:fig3}) and their connection to the so-called homoclinic snaking phenomenon~\cite{knobloch2015spatial}, followed in Subsection~\ref{tpsnsec} by a parallel study of TPs. Unless stated otherwise, the bifurcation diagrams show the quantities
\begin{align} \label{eq:phase}
    \|U\|_2=\begin{cases}
    \sqrt{\dfrac {1}{|\varOmega|} \int_\varOmega\Bra{u(\textbf{x},t){-}u_*}^2\dd \textbf{x}},& \text{for (relative) equilibria, e.g., TWs} \\ \\
    \sqrt{\dfrac {1}{\tau |\varOmega|} \int_0^{\tau} \int_\varOmega\Bra{u(\textbf{x},t){-}u_*}^2\dd \textbf{x}\dd t},& \text{for POs, including SWs and LSWs}
    \end{cases}
\end{align}
as functions of the bifurcation parameter $k_1$, and are complemented by plots of the temporal period $\tau$ (for SWs) or the speed $s$ (for TWs), as in Figs.~\ref{fig:fig3} and~\ref{f3}, respectively. For TWs stability is determined from the eigenvalues of the linearization, while SWs stability requires the computation of Floquet multipliers. Care must be taken to distinguish between instability modes in the SW invariant subspace (computed with NBCs) and those orthogonal to it, requiring PBCs. The Floquet calculations are expensive, however, and not always reliable, and so necessitate confirmation via DNS. In the figures we use thick lines to indicate stability. In particular, 
linear stability analysis of the small amplitude TW and SW states in (\ref{eq:TW_ampltds}) and (\ref{eq:SW_ampltds}) following~\cite{knobloch1986oscillatory} shows that near onset the TW are twice unstable while the SW are once unstable, a conclusion confirmed by explicit stability calculations.
\begin{figure}[ht!]
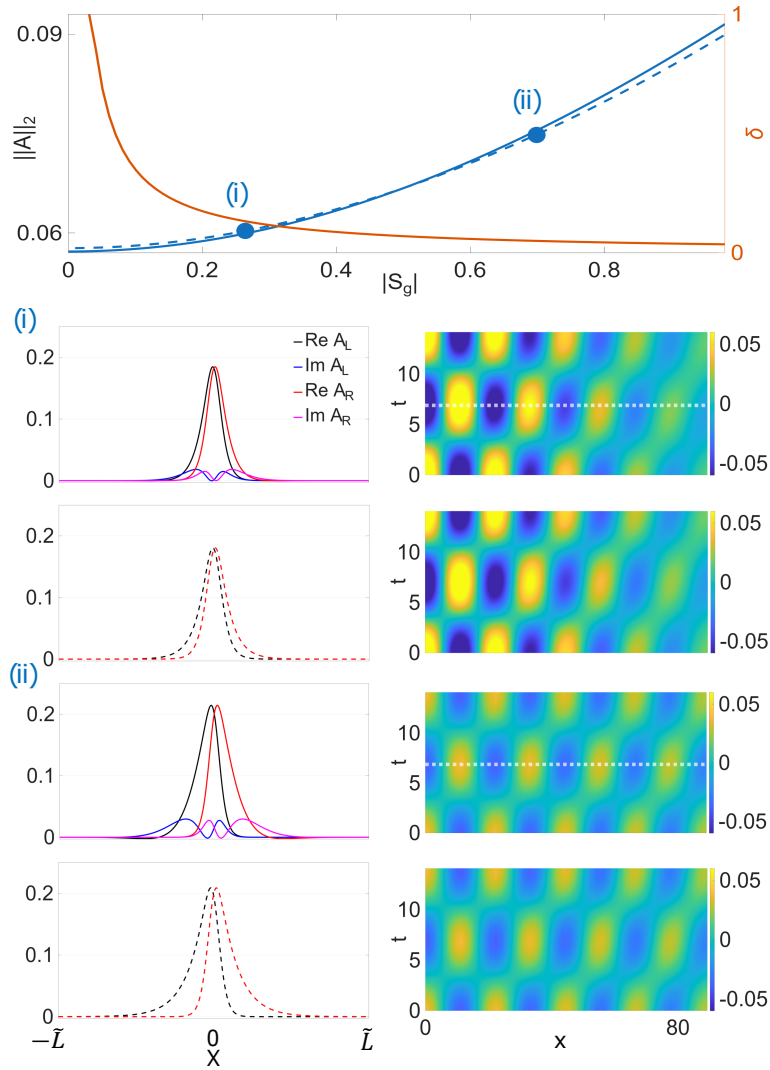

	\centering
	\ig[width=0.65\linewidth]{"Fig4-eps-converted-to"}
        \caption{Top: bifurcation diagram for LSWs of~\eqref{eq:final_ampltd} with NBCs, shown in terms of $|S_g|$ and $\|A\|_2{=}\sqrt{ (2 \tilde L)^{-1}\int[ (\Re \, A_{\rm L})^2 + (\Im \,A_{\rm L})^2 + (\Re \, A_{\rm R})^2 + (\Im \, A_{\rm R})^2]\dd X}$ together with the corresponding $\delta$ values (right axis, orange line) indicating the distance from the threshold $k_{1c}$ in~\eqref{eq:bm0}. Solid blue line corresponds to the computed values of $\alpha_i,\beta,\gamma,\eta_i\neq 0$, while the dashed blue line corresponds to $\alpha_i=\beta=\gamma=\eta_i=0$. Lower left panels depict the profiles of the real and imaginary parts of the amplitudes $A_{\rm L}, A_{\rm R}$ for (i) $S_g=0.26$, (ii) $S_g=0.7$, corresponding to the presence of a localized source (sink) at $x=0$ when $S_g>0$ ($S_g<0)$, computed on a rescaled domain with $\tilde L = \kappa L \simeq 42.19$ (see Appendix~\ref{app:amp_eq}). {Solid/dashed lines of the profiles in (i,ii), correspond to solutions at locations along the blue solid/dashed bifurcation lines in the top panel, i.e., for $\alpha_i,\beta,\gamma,\eta_i\neq 0$ and $\alpha_i,\beta,\gamma,\eta_i= 0$, respectively.} Lower right panels show the resulting space-time reconstruction of $\delta u_1 (x,t)$ over one half of the spatial domain (cf. Fig.~\ref{fig:fig3}) and a single period in time; the dotted white line is a guide to the eye showing a minor space-dependent phase shift. Continuation is conducted using $|S_g|$ as the primary parameter with $\alpha_i$ as a secondary parameter. The impact of $\alpha_i$ on $\tau$ is shown in Fig.~\ref{fig:fig3} (lower left panel, dashed brown line).
      }
	\label{fig:fig3d}
\end{figure}
\begin{figure}[tp!]
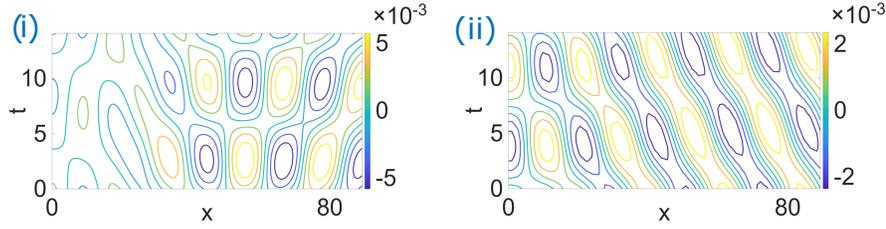

	\centering
	\ig[width=0.75\linewidth]{"Fig5contourcolor-eps-converted-to"}
        \caption{Space-time plots showing contours of the difference between $\delta u_1(x,t)$ from in~\eqref{eq:final_ampltd} obtained with complex coefficients (top right panels of (i) and (ii) in Fig.~\ref{fig:fig3d}) and that obtained with real coefficients (bottom right panels of (i) and (ii) in Fig.~\ref{fig:fig3d}, for (i) $S_g=0.26$ and (ii) $S_g=0.7$. The latter shows an unambiguous phase lag between the outer regions and the LSW center at $x=0$.
        }
	\label{fig:fig3dphase}
\end{figure}

\subsection{Homoclinic snaking of localized standing waves (LSWs)}
\label{lswsec}
We recall that LSWs are oscillating localized structures or oscillons embedded in a uniform background, here $\bU=\bU_*$. We use time-dependent numerical continuation to compute LSWs using cutoffs of numerically obtained SWs to initialize the Newton method, regardless of their stability. This process requires careful choice of the cutoff details (e.g., the spatial location of the cut), and sometimes resetting of other parameters, such as $k_1$. In the present case, the branch of LSWs in Fig.~\ref{fig:fig3} (top left panel) was obtained via an initial cutoff near $x=10$ of the small amplitude SW solution at location (ii); note that due to the NBC the midpoint of the LSWs is at $x=0$. Numerical continuation results in a snaking branch of even LSW (brown): the branch bifurcates from the SW branch (pink) at small amplitude, close to the finite wavenumber Hopf onset and as $\|U\|_2$ increases it folds over and reconnects with the SW branch near its fold. The corresponding SW and LSW temporal periods are shown in the bottom left panel of Fig.~\ref{fig:fig3}. DNS indicate that all LSWs along the (brown) snaking branch are unstable for this set of parameter values (and their vicinity, not shown). 

The behavior of the solutions along the LSW branch follows that familiar from homoclinic snaking of steady localized states in the cubic-quintic Swift-Hohenberg equation~\cite{burke2007localized}. At the lower end, the LSW states resemble a gently modulated wave packet but this state quickly localizes [profile (ii)], and after the first fold on the left, proceeds to add a half-wavelength on either side of the structure over each back-and-forth oscillation of the branch. Thus, two wavelengths have been added by location (iii) and four by (iv). Once the solutions are in contact with the right domain boundary and start resembling a "hole" in a SW state [panel (iv)], the branch turns over towards smaller $k_1$, and the hole fills in by the time the LSW branch terminates on the SW branch near its fold [panel (v)]. Since the continuation was conducted on a finite domain, the LSW branch in fact emerges from the SW branch in a secondary bifurcation at small but finite amplitude, cf.~\cite{knobloch2015spatial}. In fact, we expect two LSW branches to bifurcate from SW simultaneously, one with an even number of oscillating cells and the other with an odd number of cells. Only the former can be computed with NBCs. 

A close inspection of the LSWs reveals a space-dependent phase shift, best seen in profiles (iv) and (v) in Fig.~\ref{fig:fig3}. These profiles show that the oscillation near the boundary of the LSW state lags behind the oscillation in the middle. To determine the origin of this phase lag, we exploit an exact localized solution of~\eqref{eq:final_ampltd} with $S_g=0$~\cite{hockingstewartson1972,popp1998cubic} as an initial guess for continuation, namely $(A_L,A_R)=(A_{\text{LSW}},A_{\text{LSW}})$, where 
\begin{align}\label{eq:clSW}
    A_{\text{LSW}} = K \Lambda e^{i\Omega_{\text{LSW}} T}\text{sech}{\left(K X\right)}^{1+i \Theta}
\end{align}
and
\begin{align*}
    K &=\pm \sqrt{\frac{\alpha_r}{1-2 \beta \Theta-\Theta^2}},\quad \Lambda =\pm
    \sqrt{\frac{2 -3 \beta \Theta -\Theta^2}{1 + \eta_r}},\quad 
\Omega_{\text{LSW}} = -\alpha_i+K ^2 \left(\beta + 2 \Theta -\beta  \Theta^2\right),\\
    \Theta &=\frac{-3 \Bra{\beta\left(\gamma +\eta_i\right) +1 + \eta_r} + \sqrt{9\Bra{\beta\left(\gamma +\eta_i\right) + 1 + \eta_r}^2 + 8\Bra{\gamma + \eta_i- \beta \bra{1 + \eta_r}}^2}}{2 \Bra{\gamma + \eta_i- \beta \bra{1 + \eta_r}}},
\end{align*}
with $\alpha_r,\eta_r>0$. These LSWs bifurcate subcritically (i.e., in the same direction as the uniform SW, $A_{SW}$), in agreement with Fig.~\ref{fig:fig3}. Note that for $S_g=0$, the individual amplitudes are symmetric with respect to ${\cal R}_1:\Bra{A_{\text{L}}(X),A_{\text{R}}(X)}\to\Bra{A_{\text{L}}(-X),A_{\text{R}}(-X)}$, i.e., about the localization center $X=0$. In addition, the equations are also symmetric with respect to ${\cal R}_2:\Bra{A_{\text{L}}(X),A_{\text{R}}(X)}\to\Bra{A_{\text{R}}(-X),A_{\text{L}}(-X)}$. When $S_g\ne0$ the former symmetry is broken, while the latter is preserved.

We use~\eqref{eq:clSW} to converge to a solution of~\eqref{eq:final_ampltd} for small $|S_g|$ (effectively as a perturbation),
after which we conduct continuation in $|S_g|$; the results are shown in the top panel in Fig.~\ref{fig:fig3d}. The amplitudes $(A_{\text{L}},A_{\text{R}})$ peak on opposite sides of $X=0$, and the reconstruction of $\delta u_1(x,t)$ results in LSW in the form of a localized source (sink) if $S_g>0$ ($S_g<0$) (the former are shown in the bottom right panels in Fig.~\ref{fig:fig3d}). The space-time reconstruction plots reveal a weak space-dependent phase lag, as shown by the dotted white line at about $t=6$. To better illustrate this phase lag, we conduct continuation with real coefficients, $\alpha_i=\beta=\gamma=\eta_i=0$, for which the initial LSW guess reads
\begin{equation}\label{eq:rlSW}
A_{\text{LSW}} = \sqrt{\frac{2\alpha_r}{1 + \eta_r}}\, \text{sech}\, \left(\sqrt{\alpha_r}X\right).
\end{equation}
These continuation results are also shown in Fig.~\ref{fig:fig3d} using dashed lines; here $\alpha_i$ is also used as a secondary continuation parameter, but its value remains always numerically negligible (effectively zero). In Fig.~\ref{fig:fig3dphase}, we show the difference between $\delta u_1(x,t)$ reconstructed from~\eqref{eq:final_ampltd} with complex coefficients and the corresponding result with real coefficients, i.e., subtracting the bottom right panels from their top right counterparts in Figs.~\ref{fig:fig3d}(i) and (ii), respectively.
In Fig.~\ref{fig:fig3dphase}(i) the result is noisy near $x=0$ because of the small values involved, but panel (ii) reveals an unambiguous phase lag between the outer regions and the LSW center at $x=0$. These results indicate that the phase lags identified in Fig.~\ref{fig:fig3} far from onset are, in fact, inherited from weaker phase lags associated with the LSWs near onset and that these are amplified as the group speed $|S_g|$ increases.

\subsection{TWs, TPs, JOs, JO trains, and mixed states over a moderate domain}\label{jtsec} 

In~\cite{knobloch2021origin}, we showed that TPs in relatively small domains ($L=80$) arise via wavelength-doubling bifurcations that turn, for example, a 4TW state into a 2TW state, followed by a further instability yielding a 1TP state (see Fig.~2 of~\cite{knobloch2021origin} and Fig.~2 of the associated supplementary material). Here, we briefly revisit the TWs and the states bifurcating from them on a moderate length domain ($L=120$), roughly corresponding to the perimeter length of the disk of radius $R=20$ considered in Section~\ref{sec:2D}. 
\begin{figure}[tp!]
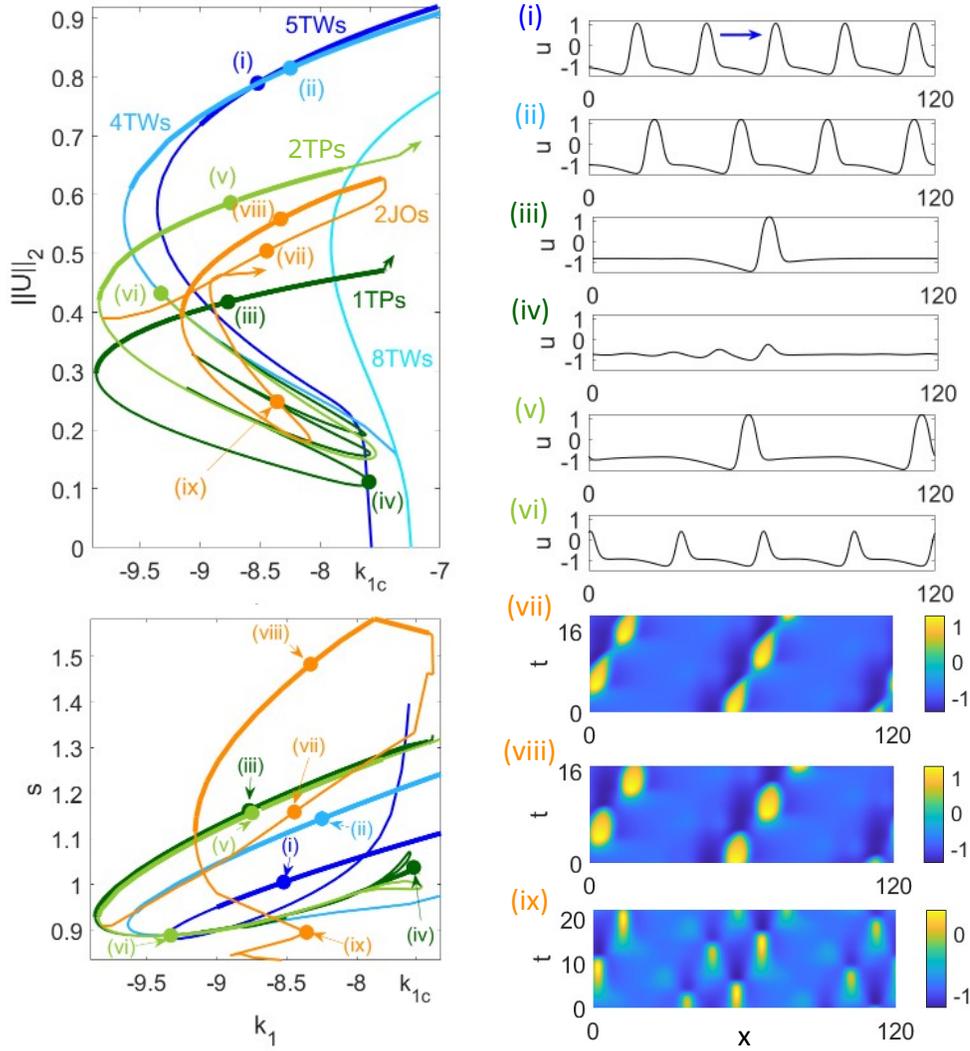

\centering
	\ig[width=0.83\linewidth]{"fig5_new-eps-converted-to"}
   \caption{Bifurcation diagram as in Fig.~\ref{fig:fig3}, showing branches of 5TW (dark blue), 8TW (light blue), and 4TW (medium blue) together with 1TP (dark green) and 2TP (light green) and a 2JO branch (orange) that bifurcates from 2TP, together with selected space-time plots at locations (i)-(ix). The space-time plots (vii)-(ix) are POs shown over two periods $\tau$ in the comoving frame. Thick (thin) lines indicate stable (unstable) states and the arrow in (i) indicates the propagation direction.}
   \label{f3}
\end{figure}
\begin{figure}[ht!]
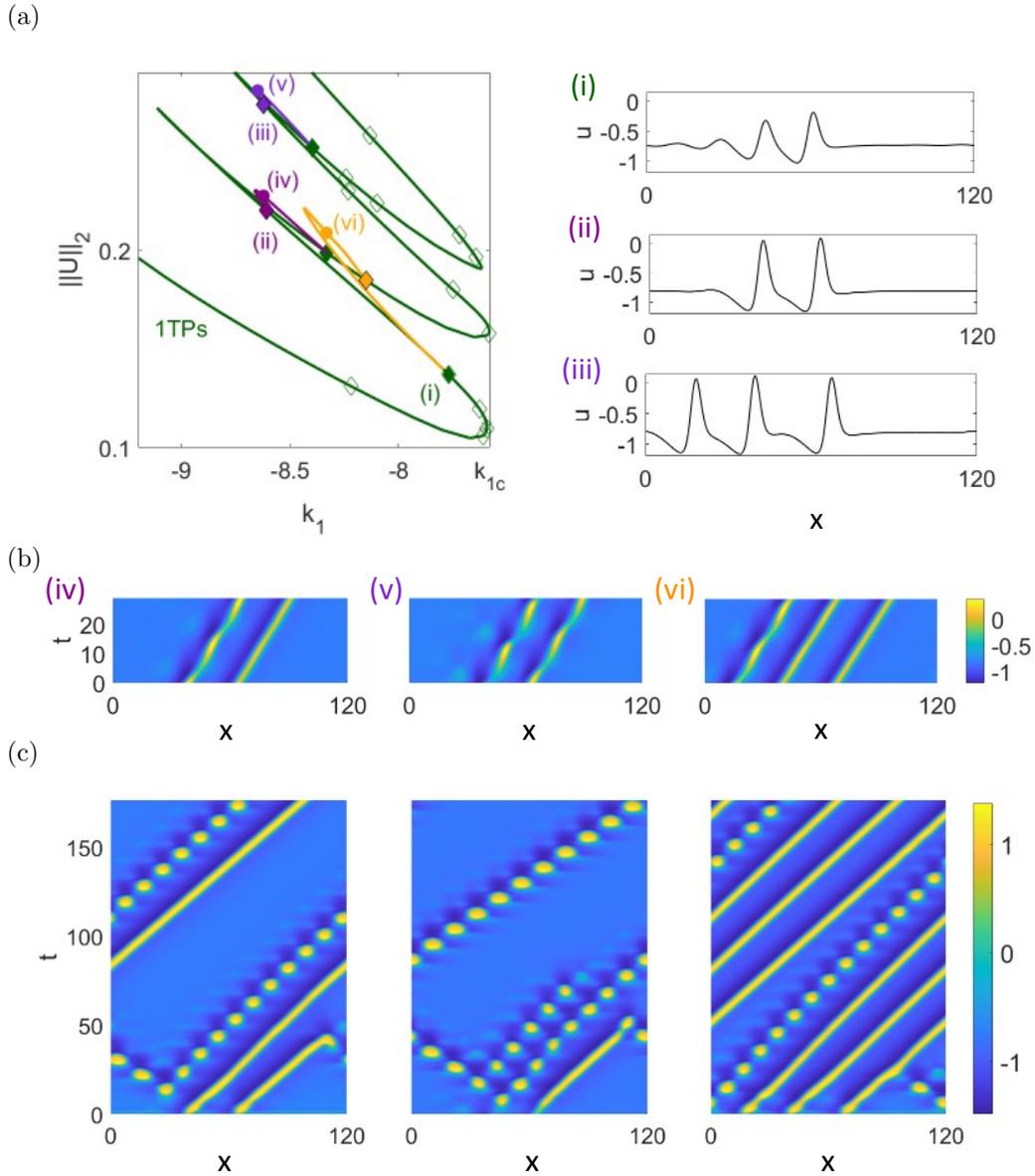

\bce
\btab{l}{{\sm (a)}\\
~ \ig[width=0.85\linewidth]{"fig6_newA-eps-converted-to"}\\[-2mm]
{\sm (b)}\\ 
~~~~\ig[width=0.85\linewidth]{"fig6_newB-eps-converted-to"}\\[-2mm]
{\sm (c)}\\
~~~\ig[width=0.85\linewidth]{"fig6_newC-eps-converted-to"}
}
\ece
   \caption{(a) Bifurcation diagram as in Fig.~\ref{f3}, focusing on homoclinic snaking region of 1TP solutions, showing a collection of Hopf bifurcations (HPs, diamonds) along with the bifurcating secondary branches of POs. (b) Selected space-time plots of three PO solutions in (a) over two temporal periods. (c) DNS at $k_1=-8.3$, initialized with the $t=0$ time slice of the POs in (b). All solutions in (a) and (b) are unstable.}
   \label{f4}
\end{figure}

In Fig.~\ref{f3} we present a bifurcation diagram for these states together with representative solutions. This bifurcation diagram is to be superposed on the bifurcation diagram in Fig.~\ref{fig:fig3}. The dark blue branch is the primary TW branch, which, for $L=120$, comprises five wavelengths (5TW), see
profile (i). The branch bifurcates subcritically and gains stability above a fold at $k_1\simeq-9.3$. We also compute a primary branch of 8TW (light blue branch) from which 4TW solutions emerge via a wavelength-doubling bifurcation [medium blue line and profile (ii)]. From solution profiles (i) or (ii), we generate single pulses by cutting off all but one pulse which we converge via DNS to stable pulse (1TP) solutions [dark green line and profile (iii)]. Continuing the 1TPs to larger $k_1$ we see that the branch loses stability in a Hopf bifurcation around $k_1\simeq-7.5$, and subsequently acquires small oscillations and spatial defects (continuation of the analogous branch on larger domains is elaborated in Fig.~\ref{fig:fig4}). 
As $k_1$ decreases, the branch undergoes a fold, where it loses stability. Continuing the branch past the fold results in unstable small amplitude traveling pulses for comparison with the larger amplitude stable 1TPs [e.g., profile (iii)], gaining spatial oscillations in the tail that grow towards the first fold on the right, near $k_1=k_{1c}$ [see profile (iv)]. The characteristics of these oscillations are discussed for larger domains in Subsection~\ref{tpsnsec}.
After the bottom right fold, the TP branch undergoes a number of alternating folds, with each back-and-forth pair resulting in the addition of an identical pulse, as shown by profiles (i) and (ii) in Fig.~\ref{f4}, forming a bound state of pulses, all traveling with a common speed. This process continues, leading to a gradual increase in the number of pulses until the available domain is filled and the TP branch terminates on the 4TW branch (not shown). This behavior is suggestive of {homoclinic snaking}, and is investigated in greater detail on a larger domain ($L=400$) in Subsection~\ref{tpsnsec}.
\begin{figure}[tp!]
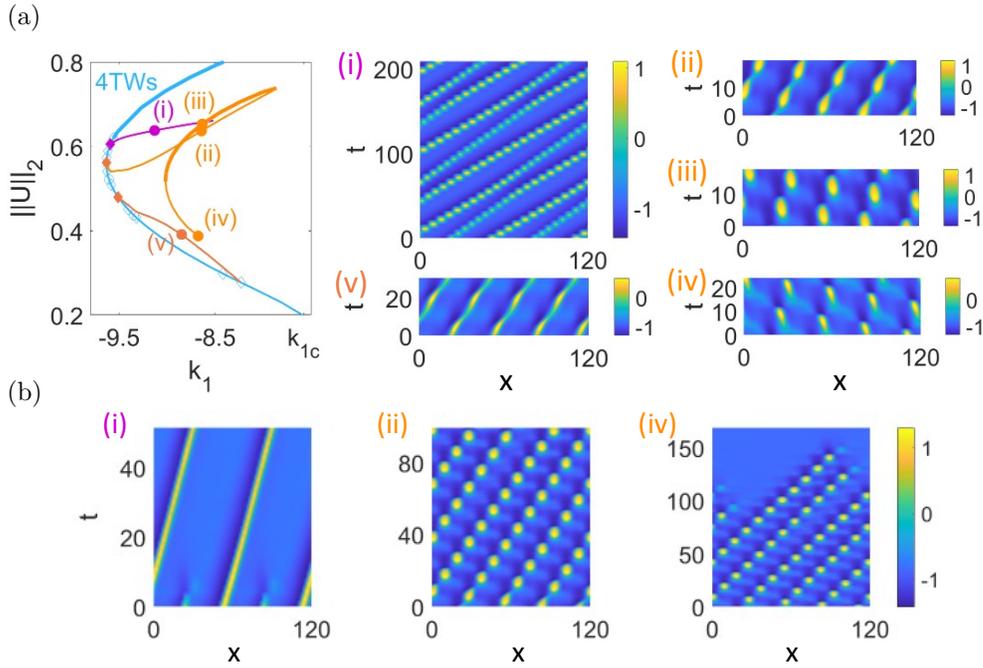

\bce
\btab{l}{{\sm (a)}\\
\ig[width=0.85\linewidth]{"fig7_newA-eps-converted-to"}\\[-4mm]
{\sm (b)}\\ \qquad
\ig[width=0.77\linewidth]{"fig7_newB-eps-converted-to"}
    }
\ece 
   \caption{(a) Bifurcation diagram as in Fig.~\ref{f4}, focusing on Hopf bifurcations (HPs, diamonds) on the 4TW branch (light blue, see also Fig.~\ref{f3}) in the vicinity of the left fold. Here, the HPs correspond to large modulation period $\tau$ leading to periodic states (i)-(v), with (i) and (v) plotted over one period, while the JO trains (ii)--(iv) are plotted over two periods, all shown in the lab frame. (b) Selected DNS at $k_1=-8.3$, initialized with the $t=0$ time slice of the corresponding periodic states in (a). Thick (thin) lines indicate stable (unstable) states.}
\label{f5}
\end{figure}

For completeness, we also compute stable 2TP states of two identical but equispaced pulses using a two-pulse cutoff of the TW profile (ii) in Fig.~\ref{f3} and subsequent DNS [profile (v)], cf.~\cite{knobloch2021origin}. The associated branch of 2TPs (light green line) recapitulates the behavior of the 1TP branch but on half of the domain and connects to the branch of 4TWs [see profile (vi)]. The large amplitude stable 1TP and 2TP states have substantially larger speeds than their unstable small amplitude snaking counterparts (see the bottom left panel in Fig.~\ref{f3}) and in both cases the speeds are almost independent of the number of pulses. As in~\cite{knobloch2021origin}, 2JO states emerge in a Hopf bifurcation below the left-most fold (orange line in Fig.~\ref{f3}), corresponding to breathing traveling states with increasingly stronger modulation as one passes from state (vii) to state (viii). The latter are stable and of larger amplitude and speed on the segment between the two folds. The branch subsequently continues to smaller amplitude and speed [see selected periodic orbit (ix)], and may also exhibit some (unstable) snaking behavior.

In fact, the 1TP and 2TP branches exhibit many HPs marked by diamonds in Fig.~\ref{f4}(a), but we focus here on some of them only (solid diamonds), i.e., those that are associated with profiles (i)-(iii). Many of the HPs in the snaking region are pairwise connected by branches of periodic orbits (POs), some of which are illustrated in Fig.~\ref{f4}(b). The figure shows (unstable) bound states of the form: (iv) 1TP-1JO (purple), (v) 1JO-1JO (dark purple) and (vi) 1JO-1TP-1TP (orange). In Fig.~\ref{f4}(c), we show DNS at $k_1=-8.3$ starting with a time--slice at $t=0$ of the (unstable) solutions in (b). For this value of $k_1$, the initial states evolve into large amplitude solutions of the form 1JO-1TP, 1JO, 3TP-1JO, respectively. We stress that the evolution is highly sensitive to the value of $k_1$ and the initial phase of the solutions in (b) chosen for the DNS; moreover, similar solutions can also be obtained from other states along the 1TP snaking structure in Fig.~\ref{f4}(a).

To further demonstrate both the richness and sensitivity of possible solutions, we discuss additional PO branches bifurcating from selectedHPs on  the 4TW branch, as shown in Fig.~\ref{f5}(a). The loss of stability of 4TWs (above the left fold) results in temporal modulation with a large period $\tau\simeq 250$ in the comoving frame traveling with speed $s\simeq 0.95$ (see Fig.~\ref{f3}). As $k_1$ increases away from this HP, the period $\tau$ decreases while $s$ slowly increases (not shown), as shown by the dark pink branch and space-time plot (i) (shown in the lab frame). These solution types are unstable, however, and DNS from the $t=0$ slice results in a 2TP state, as shown by the space-time plot (i) in panel (b). Similar branches emerge from additional HPs along the unstable
portion of the 4TW branch (not shown), while near the fold, we find POs that emerge with a much shorter period (orange line), with typical solutions shown by space-time plots (ii), (iii) and (iv). This branch undergoes a fold on the right, forming stable JO trains (iii) up to the next fold on the left; DNS from the unstable
solution (ii) in (a) converges to a JO train, as shown in (ii) in panel (b). After the left fold, the modulation decreases in amplitude, as shown in (iv) in panel (a) and DNS typically show decay to the uniform state $\bU_*$, as shown in (iv) in panel (b). Below the 4TW fold, the remaining HPs regain pairwise connections by ``short'' branches, much as in Fig.~\ref{f4}. The solutions on these branches can be either in phase [see space-time plot (v) in (a)], or exhibit various phase gradients, as shown in (v) in Fig.~\ref{f4}(b). However, DNS show that these (unstable) states typically decay to ${\bf U}_*$. We mention that similar Hopf bifurcations also occur near left folds on other nTW
branches ($n=3,5,6,\ldots$) and generate nJO-trains by the same mechanism.

To summarize, Figs.~\ref{f3}--\ref{f5} show the subcritical regime of $k_1<k_{1c}$ is filled with an increasing variety of stable and unstable states comprising TWs, TPs, JOs, JO trains and mixtures thereof, even for moderate domain sizes (here $L=120$), and this variety is expected to grow with increasing $L$. The following section is devoted to exploring one specific consequence of increasing $L$: homoclinic snaking of TPs.

\subsection{Homoclinic snaking of traveling pulses (TPs)}\label{tpsnsec}
\begin{figure}[tp!]
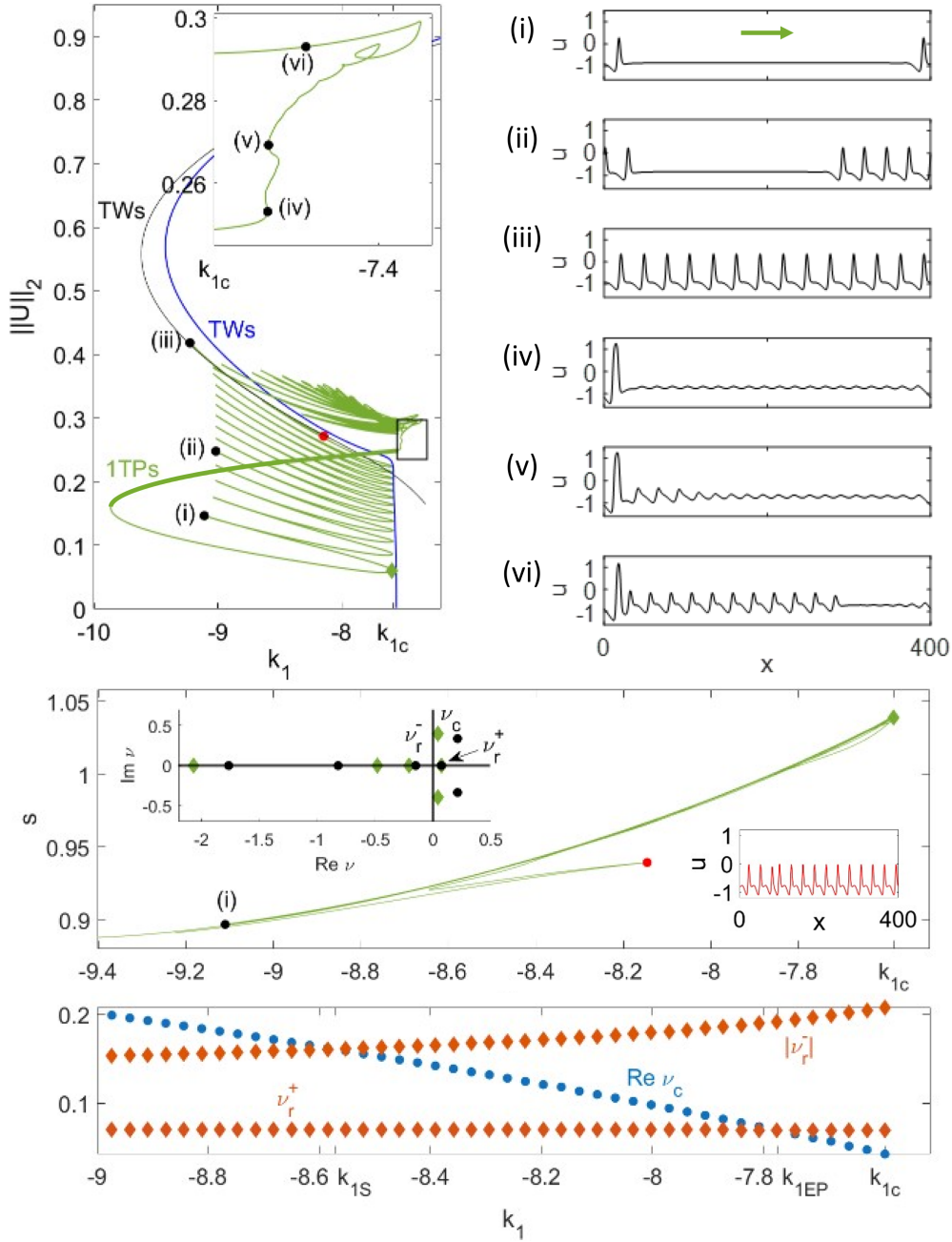

	\centering
	\ig[width=0.87\linewidth]{"fig4a2-eps-converted-to"}
	\caption{Bifurcation diagram obtained via continuation in AUTO on a periodic domain of length $L=400$ (top left panel), showing branches of uniform traveling waves with different wavelengths $\lambda$ (TWs, blue for $\lambda=25$, black for $\lambda\simeq 29.5$), together with those of traveling pulses (1TPs, green) and traveling pulse trains (TPTs, green) and sample TPT profiles (i)-(vi) shown on the right; thick (green) line indicates linear stability and the arrow in (i) indicates the propagation direction. The blue TWs bifurcate from $\bU_*$ when it first loses stability; the black TW curve corresponds to a periodic solution (iii) that emerges from a wavelength doubling bifurcation on another primary TW branch (not shown, cf.~\cite{knobloch2021origin}). The middle panel shows the speed $s$ in the snaking regime, while the left inset depicts the six spatial eigenvalues [$\nu$, see~\eqref{eq:lin_spatial}] of $\bU_*$ at the indicated locations, i.e., at the corresponding $k_1$ and $s$ values; the right inset shows the profile at the red dot prior to reaching the periodic state (iii). The bottom panel shows the location of $k_1=k_{1S}\simeq -8.57$, where the Shil'nikov ratio $\delta_S(s)\equiv|\nu^-_r|/\Re\, \nu_c=1$, where $s$ is the speed of the small amplitude 1TPs in the comoving frame (middle panel). This point is located in the snaking region and corresponds to the crossing of the real part of the complex eigenvalues ($\nu_c$, blue dots) and the absolute value of the leading negative eigenvalue ($\nu^-_r$, red diamonds) as shown in the bottom panel. Both depend on the 1TP speed $s$.}
	\label{fig:fig4}
\end{figure}

To explore the properties of the TP snaking behavior described in Fig.~\ref{f3} in greater detail, we compute and show in Fig.~\ref{fig:fig4} the corresponding results for $L=400$, obtained {in this subsection} using the package AUTO~\cite{doedel2012auto} employing periodic boundary conditions. We again start from a stable 1TP solution, here at $k_1=-8.2$, and continue this solution in both directions in $k_1$, resulting in the complex homoclinic snaking structure shown in the top left panel in Fig.~\ref{fig:fig4}. As before only the single TP solutions (1TPs) are linearly stable (as indicated by the thick line), and as $k_1$ decreases, the 1TP branch undergoes a fold where it loses stability. Past the fold we find the gradual appearance of small amplitude spatial oscillations in the tail of the TP profile, much as for $L=120$. Between the subsequent right and left folds the bump adjacent to the primary peak develops into a second peak while the remaining oscillations are suppressed, resulting in the 2TP state (i). This process continues, yielding a gradual increase in the number of peaks (ii), until the domain is filled (iii) and the TP branch terminates on a subsidiary TW branch (black) that arises in a wavelength-doubling bifurcation from a primary TW branch with wavenumber $q\ne q_c$ (not shown). As in Fig.~\ref{f3}, the speed $s$ of these traveling pulse groups or bound states depends only weakly on the number of pulses in the group. 

We associate this snaking behavior with the presence of complex spatial eigenvalues of $\bU_*$, in the comoving frame~\cite{knobloch2021stationary,li2025traveling}, $\xi\equiv x-st$:
\begin{equation}\label{eq:lin_spatial}
\left( {\begin{array}{*{20}c}
	{u}  \\
	{\tilde u}  \\
	{v}  \\
    {\tilde v} \\
	{w}  \\
	{\tilde w}  
	\end{array}} \right)-\left( {\begin{array}{*{20}c}
	{u_*}  \\
	{0}  \\
	{v_*}  \\
    {0}  \\
	{w_*}  \\
	{0}
	\end{array}} \right) \propto e^{\nu \xi},
\end{equation}
where $\tilde u=u_\xi$, $\tilde v=v_\xi$, and $\tilde w=w_\xi$. The inset in the middle panel of Fig.~\ref{fig:fig4} shows that of the six eigenvalues $\nu$, three are negative and three positive. Consequently, the 'stable' and 'unstable' manifolds of $\bU_*$ are both three-dimensional. The intersection of such manifolds in six dimensions is generically of codimension one, implying that each TP state is associated with a unique speed $s(k_1)$. The inset, moreover, shows that the leading `unstable' eigenvalues at the green diamond are complex while the leading `stable' eigenvalue is real and more negative, suggesting that at this location the profile of the pulse at the leading edge is monotonic and steep, while its trailing profile is gentler and oscillatory, i.e. $\bU_*$ is a saddle-focus in the comoving frame with the Shil'nikov ratio $\delta_S(s)\equiv |\nu^-_r|/\Re\, \nu_c>1$. Under these conditions, we expect locking of widely separated pulses, cf.~\cite{li2025traveling}, i.e., we expect bound states of pulses for $k_{1S}<k_1<k_{1c}$, where $k_{1S}\simeq -8.57$ (Fig.~\ref{fig:fig4}). It is not clear why the snaking TP branch extends below this $k_1$ value, although the detailed study of a different nonreversible system~\cite{nonreversible,champneys2009unfolding} may indicate the way forward.

Indeed, the leading `unstable' eigenvalue at location (i) is real, albeit still smaller than the modulus of the leading `stable' eigenvalue, indicating the presence of an exchange point $\nu^+_r=\Re\, \nu_c$ at $k_1=k_{1EP}\simeq -7.78$, where $k_{1S}<k_{1EP}<k_{1c}$, that results in the elimination of the small tail oscillations far from each pulse. In systems with left-right symmetry, the passage of $k_1$ through $k_1=k_{1EP}$ results in pulse-pulse repulsion~\cite{knobloch2021stationary} and hence the destruction of bound states of two or more pulses. This is not the case here, where the pulses are strongly asymmetric, and oscillations at the trailing edge may be present over a large but ultimately finite distance, indicating the presence of bound states with small to moderate separation until $\delta_S(s)$ reaches $\delta_S(s)=1$, i.e., in the larger interval $k_{1S}<k_1<k_{1c}$.

We make two further observations of interest. In contrast to traveling structures in the Swift-Hohenberg equation with third order dispersion~\cite{burke2009swift}, here the pulse groups are organized in a continuous snaking structure instead of a stack of isolas, cf.~\cite{beck2009snakes}. Moreover, traveling pulse train (TPT) states appear to accumulate on the primary TW branch (blue), much as found in~\cite{lojacono2017}. Figure~\ref{fig:fig4} shows that this behavior is associated with canard-like behavior of the primary TW branch, cf.~Ref.~\cite{vo2025canards}.

Once the domain is filled with pulses the snaking process ends and the TP branch terminates on a TW branch (iii). This branch can also be followed as a function of $k_1$ and is shown in Fig.~\ref{fig:fig4} (black). Note that just before the domain is filled, the TPT branch undergoes a large excursion (red dot) in which both the spatial period and speed readjust (see the profile in the right inset of the bottom panel) prior to connecting to the TW branch at location (iii). The latter is distinct from the TW branch that bifurcates subcritically from $\bU_*$ at $k_{1c}$ with wavelength $\lambda=400/16=25$, a wavelength that differs from the wavelength $\lambda=400/14\simeq 28.6$ of the TW state (iii); as already mentioned the latter bifurcates from a TW state with wavelength $L=400/28\simeq 14.3$ that can be traced to a bifurcation point on $\bU_*$ that lies in $k_1>k_{1c}$.

Extending the 1TP branch in the opposite direction, past $k_{1c}$ and toward increasing $\|U\|_2$ (top panel, inset), also results in a complex solution structure comprising different wavelengths and defect structures. Sample solution profiles are shown in panels (iv)-(vi) and resemble those shown in~\cite{parra}. i.e., a pulse embedded in a background oscillation. 

\section{Rotating and oscillating spots on a disk}\label{sec:2D}

We now extend our 1D results to patterns on disks, aiming to identify both qualitative similarities and qualitative differences with the 1D case, with the Neumann boundary conditions (NBCs) 
\begin{equation}
({\bf \widehat n}\cdot\nabla) {\bf U}|_{\partial\varOmega}=0,
\end{equation}
where ${\bf \widehat n}$ is the unit normal. In 2D, we find it convenient to distinguish between solutions attached to the disk boundary and those present in the disk bulk. We use the variable $\xi\in [-\pi R,\pi R)$ to describe the former and the cut $x\in (-R,R)$ at $y=0$ to describe the latter and refer to the supplementary movies for the corresponding space--time plots. For numerical convenience, all computations are performed with $D_w=50$ instead of $D_w=100$ as used in Section~\ref{sec:1D}. {The results demonstrate not only the complexity of the emerging patterns but also their sensitivity to initial conditions arising from complex transients present in this geometry.}

In general, we follow the strategy employed in Section~\ref{sec:1D}. We first identify the analog of the finite wavenumber Hopf bifurcation in 1D, here corresponding to a nonzero azimuthal wavenumber $m$. The linearized problem around $\bU_*$ on the disk supports two classes of eigenfunctions, bulk and wall states given in terms of Bessel functions 
\cite{goldsteinJFM1993,verschueren2021localized}, and both can undergo a wave instability. We compute the resulting branches using numerical continuation and seek secondary bifurcations or use cutoffs of periodic solutions to generate an initial guess for the continuation of azimuthally localized structures. The continuation results are supplemented with DNS to identify states we cannot obtain via numerical continuation and their stability. To present, in our results, we use different hues of blue (pink) for rotating (oscillating) azimuthally periodic states, and green for localized states. Oscillations with $m=0$ correspond to axisymmetric oscillations in the bulk referred to as target waves, of which we plot just one example, in Fig.~\ref{fig:fig7}. 
\begin{figure}[tp!]
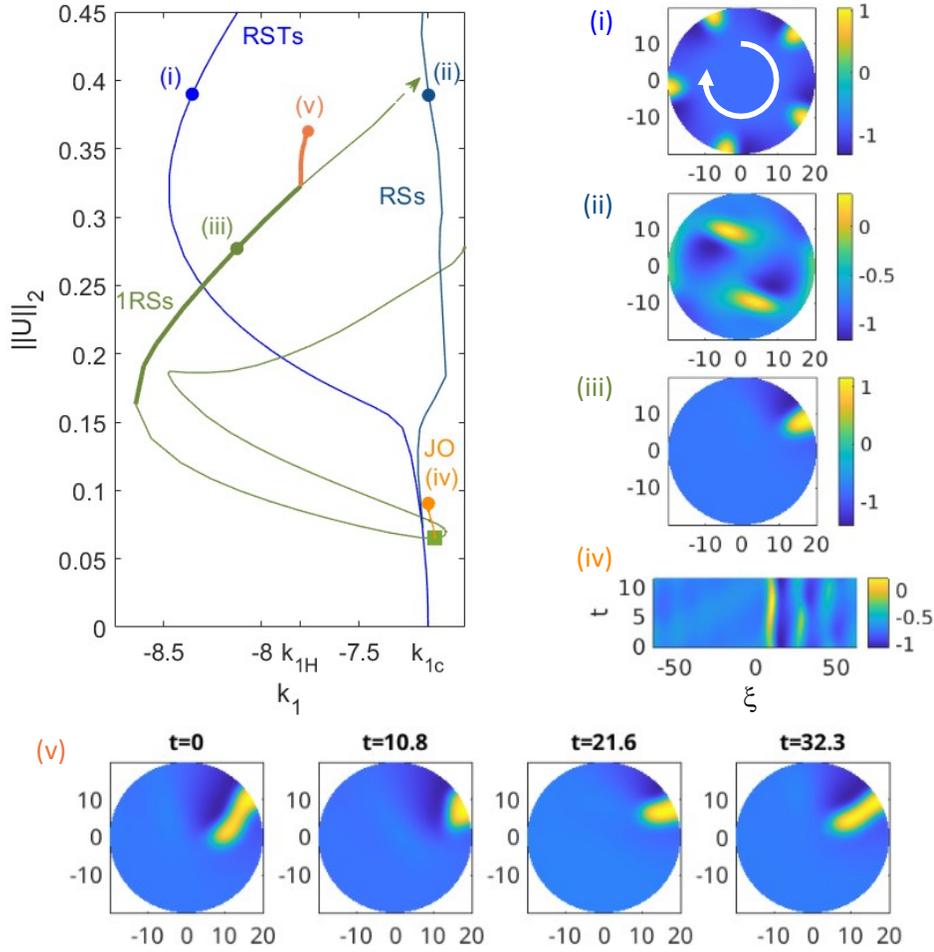

	\centering
	\ig[width=0.8\linewidth]{"fig5B-eps-converted-to"}\\
    \caption{Bifurcation diagram obtained via continuation on a disk of radius $R=20$ with $D_w=50$, showing branches of wall-attached rotating spot trains [RSTs, blue, state (i) at $k_1=-8.354$], rotating spots [1RSs, green, state (iii) at $k_1=-8.125$] and two short PO segments: JO [orange, state (iv) at $k_1=-7.148$] and a pulsating filament [also orange, state (v) at $k_1=-7.762$, in corotating frame]. Bulk rotating spots set in very close to $k_{1c}$ [2RSs, blue-green, state (ii) at $k_1=-7.147$]. Thick lines (green and orange) indicate linear stability: the 1RS state is stable between the fold on the left and a Hopf bifurcation at $k_1\equiv k_{1H}\simeq -7.798$ and transfers stability to the filament state (v) at $k_1\equiv k_{1H}$. All states rotate in a clockwise direction as indicated in (i). In (iv), we show the space-time evolution along the disk perimeter ($\xi\in [-\pi R,\pi R)$) of a modulated 1RS state in the corotating frame arising from a Hopf bifurcation of the 1RS state (green square).}
	\label{fig:fig5}
\end{figure}
\begin{figure}[tp!]
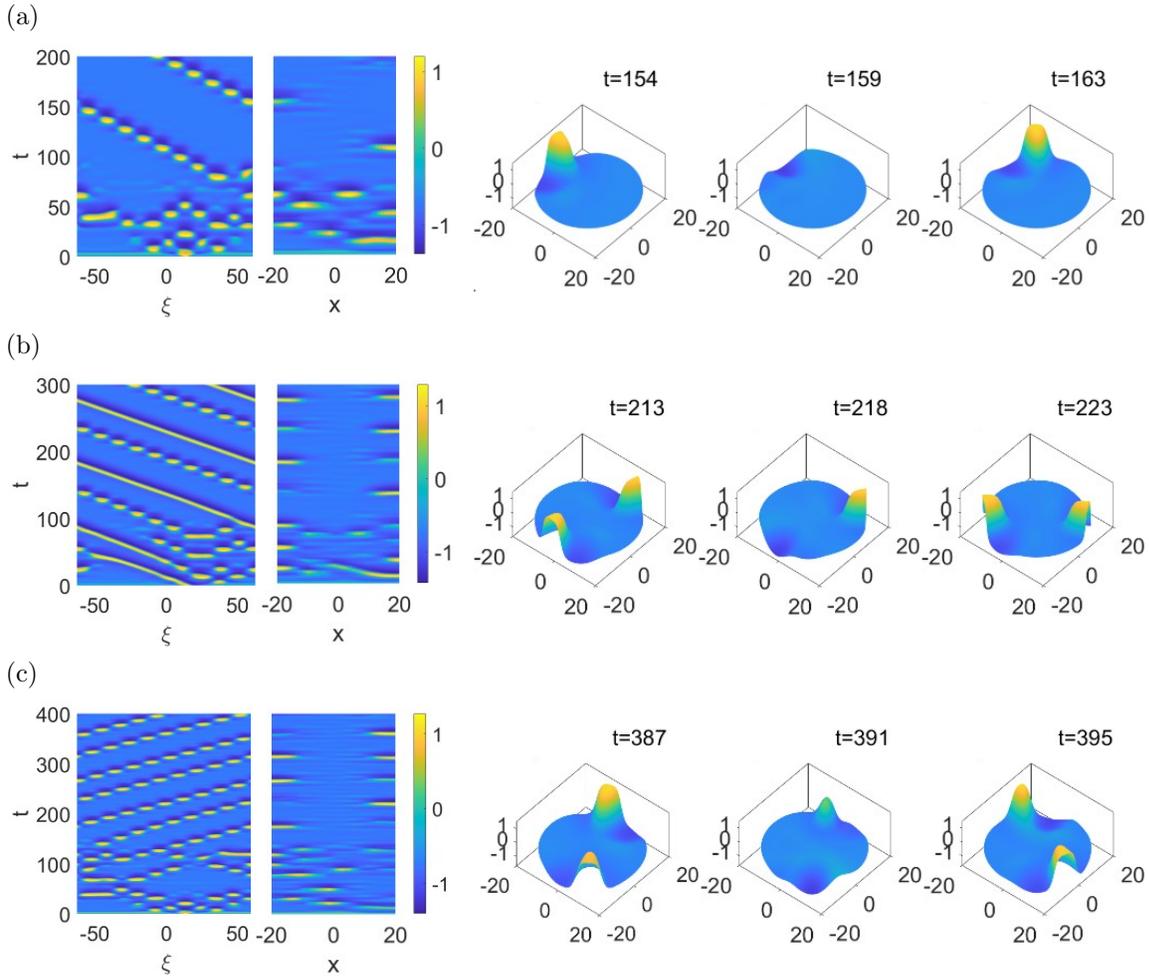

	\centering    
\btab{l}{
        {\sm (a)}\\ \ig[width=0.97\linewidth]{"fig8a-eps-converted-to"}\\
        {\sm (b)}\\ \ig[width=0.97\linewidth]{"fig8b-eps-converted-to"}\\
        {\sm (c)}\\ \ig[width=0.97\linewidth]{"fig8c-eps-converted-to"}
        }
	\caption{Space-time plots along the perimeter 
($\xi\in [-\pi R,\pi R)$) and diameter ($x\in [-R,R]$, $y=0$) together with solution snapshots at the indicated times, obtained via DNS on a disk with $R=20$, showing the evolution into (a) a jumping oscillon (JO) at $k_1=-7.8$ initialized with an oscillating spot (OS) at the boundary, (b) a RS-JO bound pair at $k_1=-7.85<k_{1H}\simeq-7.798$ initialized with a rotating spot (RS) at the boundary, and (c) a bound pair of JOs at $k_1=-7.75>k_{1H}$ initialized with an oscillating spot (OS) at the boundary. These solutions are analogs of the respective 1D states, cf.~Figs.~\ref{fig:fig1} and \ref{f4}. See supplementary \textit{movieFigure11a}, \textit{movieFigure11b}, and \textit{movieFigure11c}, respectively.}
	\label{fig:fig6}
\end{figure}

We start with a relatively small disk of radius $R=20$ but one that allows comparison with the 1D results since its perimeter is of length $\ell=40\pi\simeq 125$ and so comparable to the 1D domain used in Figs.~\ref{f3} and \ref{f4}. In Fig.~\ref{fig:fig5}, we show a partial bifurcation diagram exhibiting two primary branches of rotating spots (RSs) that bifurcate in close succession from the trivial state. The first bifurcates at $k_1=k_{1c}\simeq -7.149$ and corresponds to an $m=5$ wall mode: a snapshot of a rotating finite-amplitude wall-attached spot train (RST) is shown in profile (i). This bifurcation is followed by an $m=2$ bulk rotating spot (RS) state at $k_1=-7.146$. Both branches are subcritical, implying instability.

The branch of wall-attached RSTs (blue) contains five wavelengths (spots), and bifurcates from the finite wavenumber Hopf onset (these states are an analog of the TW state in 1D) while the bulk RS states (blue-green branch) consist of two spots. Using a cutoff of the RSTs and replacing it by $\bU_*$ for $\phi\not\in(-0.22\pi, 0.35\pi)$, we obtain an apparently disconnected branch (green) of single wall-attached RS (1RS) states, see profile (iii), together with their stability properties. In particular, the 1RS branch loses stability with increasing amplitude in a supercritical Hopf bifurcation at $k_1=k_{1H}\simeq-7.798$ (Fig.~\ref{fig:fig5}, top left panel) where stability is transferred to a branch of relative POs (orange) resembling a pulsating filament (Fig.~\ref{fig:fig5}, state (v)). On the other hand, with decreasing amplitude, the 1RS branch exhibits behavior resembling snaking, as in 1D. We find, in addition, a second branch of relative POs consisting of slowly drifting temporally oscillating wall-attached states resembling JOs (also orange), resulting from a secondary Hopf bifurcation of the 1RS state near its right fold. These wall-attached JOs are the analogs of JOs in 1D and are unstable. We show a space-time plot of this state in Fig.~\ref{fig:fig5}, state (iv). As with the other PO branch, we were able to follow only a small portion of this branch. 
\begin{figure}[tp!]
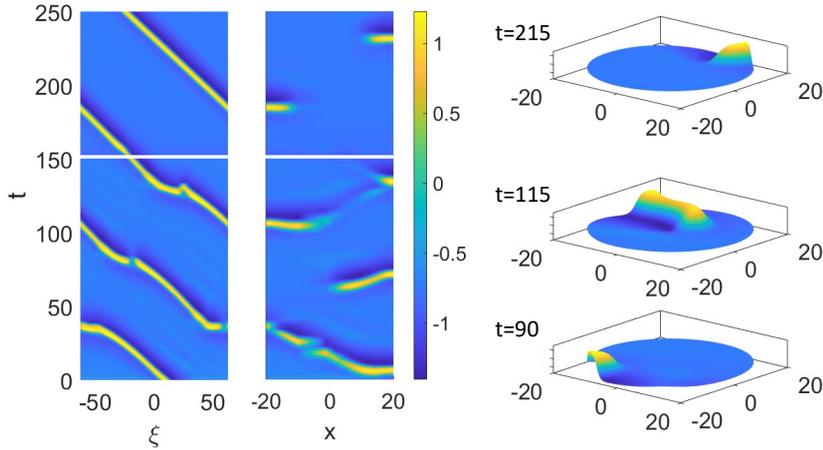

	\centering
        \ig[width=0.7\linewidth]{"fig8d-eps-converted-to"}\\
        \caption{Space-time plots along the perimeter ($\xi{\in}[-\pi R,\pi R]$, left panels) and diameter ($x{\in}[-R,R]$, $y{=}0$, right panels) obtained via DNS on a disk with $R{=}20$, initialized with a single traveling spot at the boundary, showing the resulting evolution. The white line marks the change from $k_1{=}-7.6{>}k_{1H}\simeq-7.798$ to $k_1{=}-7.9<k_{1H}$. See supplementary \textit{movieFigure12}. 
        }
        \label{fig:fig6b}
\end{figure}
\begin{figure}[ht!]
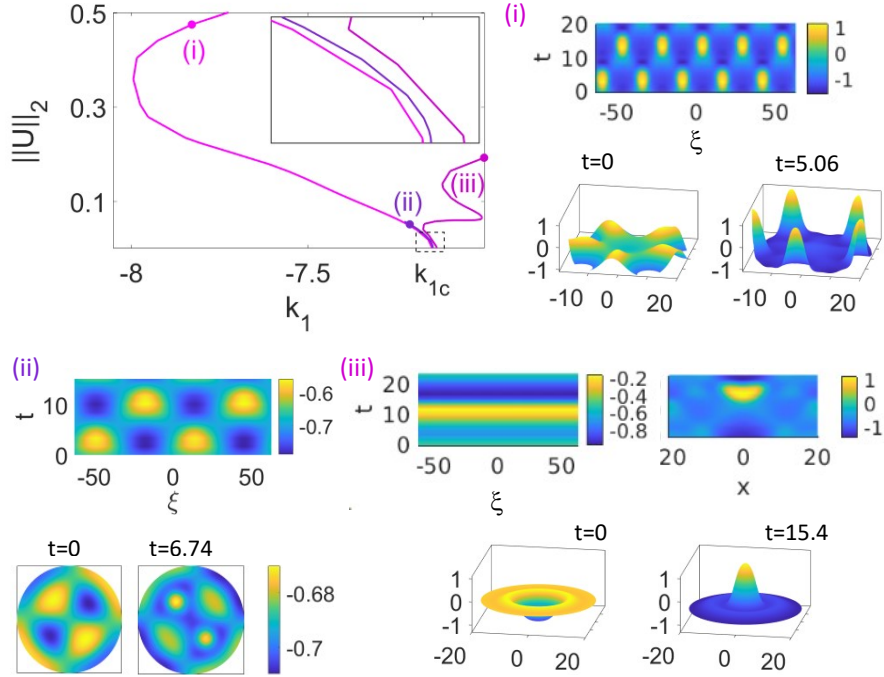

	\centering
	\ig[width=0.75\linewidth]{"fig6-eps-converted-to"}
\vskip 2mm
	\caption{Bifurcation diagram showing different branches of standing oscillations, both wall-attached [state (i), $k_1=-7.828$] and bulk [state (ii), $k_1=-7.21$], obtained via continuation on a disk of radius $R=20$, together with space-time plots and solution snapshots of wall-localized oscillations in terms of both the disk perimeter ($\xi\in [-\pi R,\pi R]$) and the disk diameter ($x\in [-R,R]$, $y=0$). Panel (iii) shows an axisymmetric target state at $k_1=-7.001$.}
	\label{fig:fig7}
\end{figure}

A clearer analog of 1D JOs is shown in Fig.~\ref{fig:fig6}(a). This state was obtained via DNS starting from the 1RS profile at location (iii) in Fig.~\ref{fig:fig5} upon changing $k_1$ to $k_1=-7.8$, {and not from the pulsating filament state (v) that emanates from $k_{1H}$. The wall-attached JO (Fig.~\ref{fig:fig6}(a)) may connect to state (iv) in Fig.~\ref{fig:fig5} but we have been unable to confirm this conjecture owing to convergence problems.}
In Figs.~\ref{fig:fig6}(b,c), we demonstrate the presence of additional states, a locked RS-JO pair and a locked JO pair, for comparison with the 1D solutions shown in Figs.~\ref{fig:fig1} and~\ref{f4}. As in 1D, the origin of these remains to be established.

Figure~\ref{fig:fig6b} shows a two-frequency filament state found through DNS initialized with a single spot on the boundary. After a complex transient that involves both wall and bulk oscillations the system settles into a state resembling a one-armed spiral. The figure shows a stable steadily-rotating filament at $k_1=-7.9<k_{1H}$ that extends from the boundary partways into the bulk. However, at $k_1$ values closer to $k_1=k_{1H}\simeq-7.798$ this state destabilizes and acquires a second frequency. The supplementary movie shows that the filament is rather stiff and propagates in both the azimuthal and radial directions, resulting in a broadside collision with the wall. This collision annihilates most of the filament, but leaves a spot at the back that serves to regrow it. This process repeats periodically, with the filament undergoing an incommensurate rotation in orientation between successive collisions. This state is illustrated in the initial part of the space-time plot, corresponding to $k_1=-7.6$. It is likely that this state lies on the orange branch in Fig.~\ref{fig:fig5} beyond state (v) but this has not been verified.

Next, we compute the bifurcation diagram of standing oscillations, referred as oscillating spots (OSs), shown in the top-left panel of Fig.~\ref{fig:fig7}. As in the case of the RSs, the OSs are also of two types: wall and bulk OSs. The wall OSs bifurcate from the instability onset at $k_1{=}k_{1c}$ and are also characterized by an $m{=}5$ wavenumber resulting in five equally spaced spots that oscillate synchronously and with identical amplitude, as shown in the space-time plot (i) and selected snapshots. The space-time plot (ii) shows the $m{=}2$ bulk OS state. This state bifurcates at the same $k_1$ value, $k_1=-7.146$, as the {2RS state} in Fig.~\ref{fig:fig5}, and generates a SW in the bulk between two pairs of oscillating spots. A (stable) state of this type has been seen in other reaction-diffusion systems as well \cite{chen2024}. The bifurcation diagram also shows a third branch, one that bifurcates at $k_1{=}-7.135$ and corresponds to axisymmetric oscillations with $m{=}0$ [panel (iii)]. This state consists of an oscillating pace-maker spot in the center of the disk accompanied by a weaker out-of-phase oscillation at the wall. 
\begin{figure}[tp!]
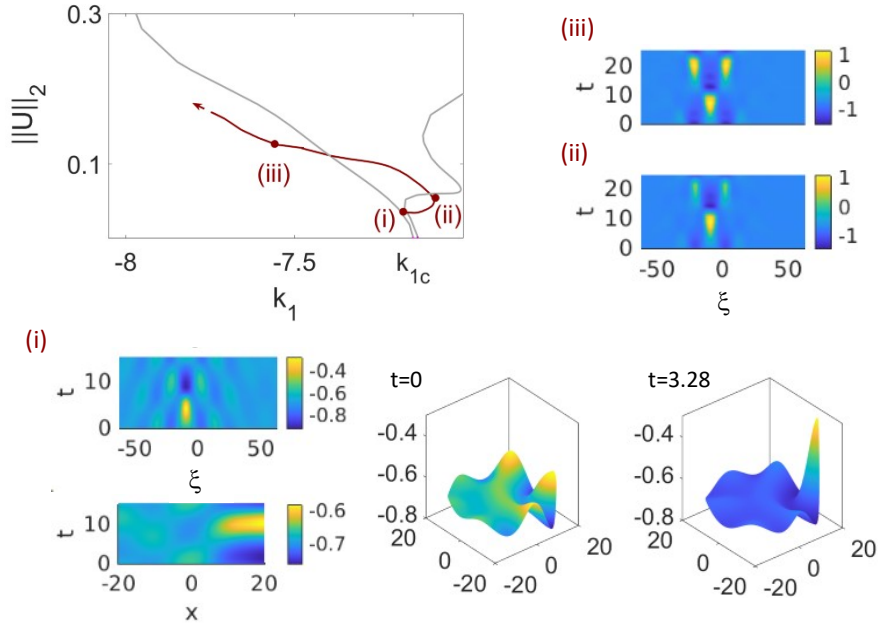

	\centering
	\ig[width=0.75\linewidth]{"fig6b-eps-converted-to"}
        \caption{As in Fig.~\ref{fig:fig7} but showing a disconnected branch of azimuthally localized spots (OSs). (i) $k_1=-7.187$, (ii) $k_1=-7.083$ and (iii) $k_1=-7.559$.}        
	\label{fig:fig7b}
\end{figure}
\begin{figure}[tp!]
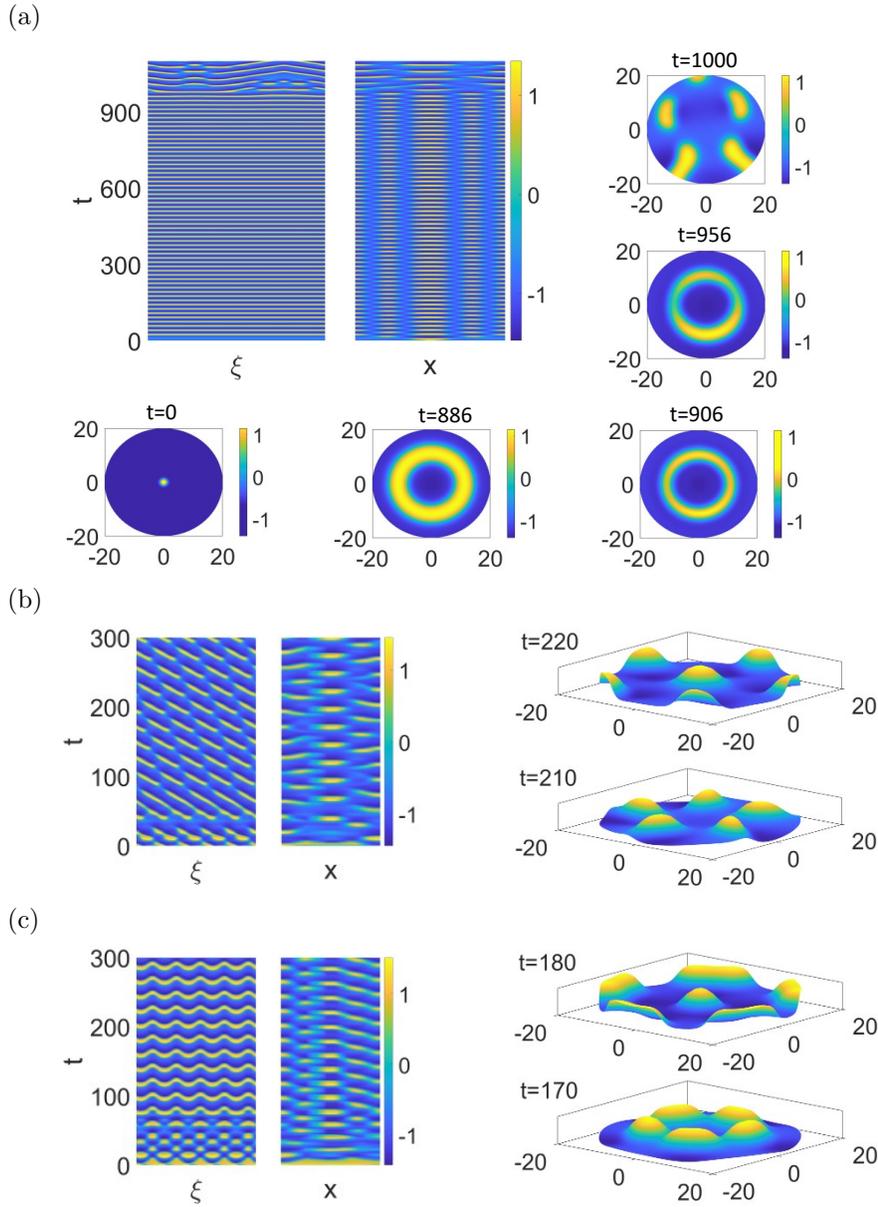

	\centering
    \btab{l}{
        {\sm (a)}\\ \hspace{0.3in}\ig[width=0.64\linewidth]{"fig13-eps-converted-to"} \\ 
        {\sm (b)}\\ \hspace{0.2in} \ig[width=0.7\linewidth]{"fig9a-eps-converted-to"}\\ 
        {\sm (c)}\\ \hspace{0.2in} \ig[width=0.7\linewidth]{"fig9b-eps-converted-to"}
        }
        \caption{Space-time plots showing time-dependent states along the perimeter ($\xi\in [-\pi R,\pi R)$) and diameter ($x\in [-R,R]$, $y=0$) obtained via DNS on a disk of radius $R=20$: (a) breakup of an axisymmetric target state at $k_{1}=-7.2<k_{1c}\simeq -7.149$ initialized with a localized spot at the bulk center, (b) rotating $m=5$ OS state at $k_{1}=-7.5<k_{1c}$ ($\bU_*$ linearly stable) initialized with periodic spots at the perimeter, and (c) standing $m=5$ OS state at $k_{1}=-6>k_{1c}$ ($\bU_*$ linearly unstable) initialized in the same way as in (b). See supplementary \textit{movieFigure15a}, \textit{movieFigure15b}, and \textit{movieFigure15c}, respectively.}
	\label{fig:fig8}
\end{figure}

We also identified, and show in Fig.~\ref{fig:fig7b}, a (disconnected) branch of azimuthally localized wall spots, which we see as analogs of the LSWs in Fig.~\ref{fig:fig3}. To obtain such isolated OSs we used a cutoff of a wall-attached SW as an initial guess for steady state continuation and obtained a branch (brown) of wall-attached OSs [Fig.~\ref{fig:fig7b}, panels (i)-(iii)]. Further continuation of these OSs (to $k_1<-8$, say) proved impossible owing to convergence issues; in particular, it is not known if the computed OS segment is part of an OS snaking branch. 

While the continuation of oscillating solution branches on a disk is numerically challenging, the computed wall-attached states, both rotating and standing, demonstrate considerable similarity to the 1D results. {We expect (and have seen) that with increasing forcing $k_1$ these states invade the disk bulk, as in the case of steady wall-attached states on a disk described by the real Swift-Hohenberg equation~\cite{verschueren2021localized}.} We have, in addition, computed rotating spot states confined entirely to the bulk, using both numerical continuation and DNS with spot-like initial conditions in the bulk and at the perimeter, resulting in mixed RS and OS states in the subcritical regime ($k_1<k_{1c}$) and standing OSs in the supercritical regime ($k_1>k_{1c}$), as shown in Fig.~\ref{fig:fig8}.
\begin{figure}[tp!]
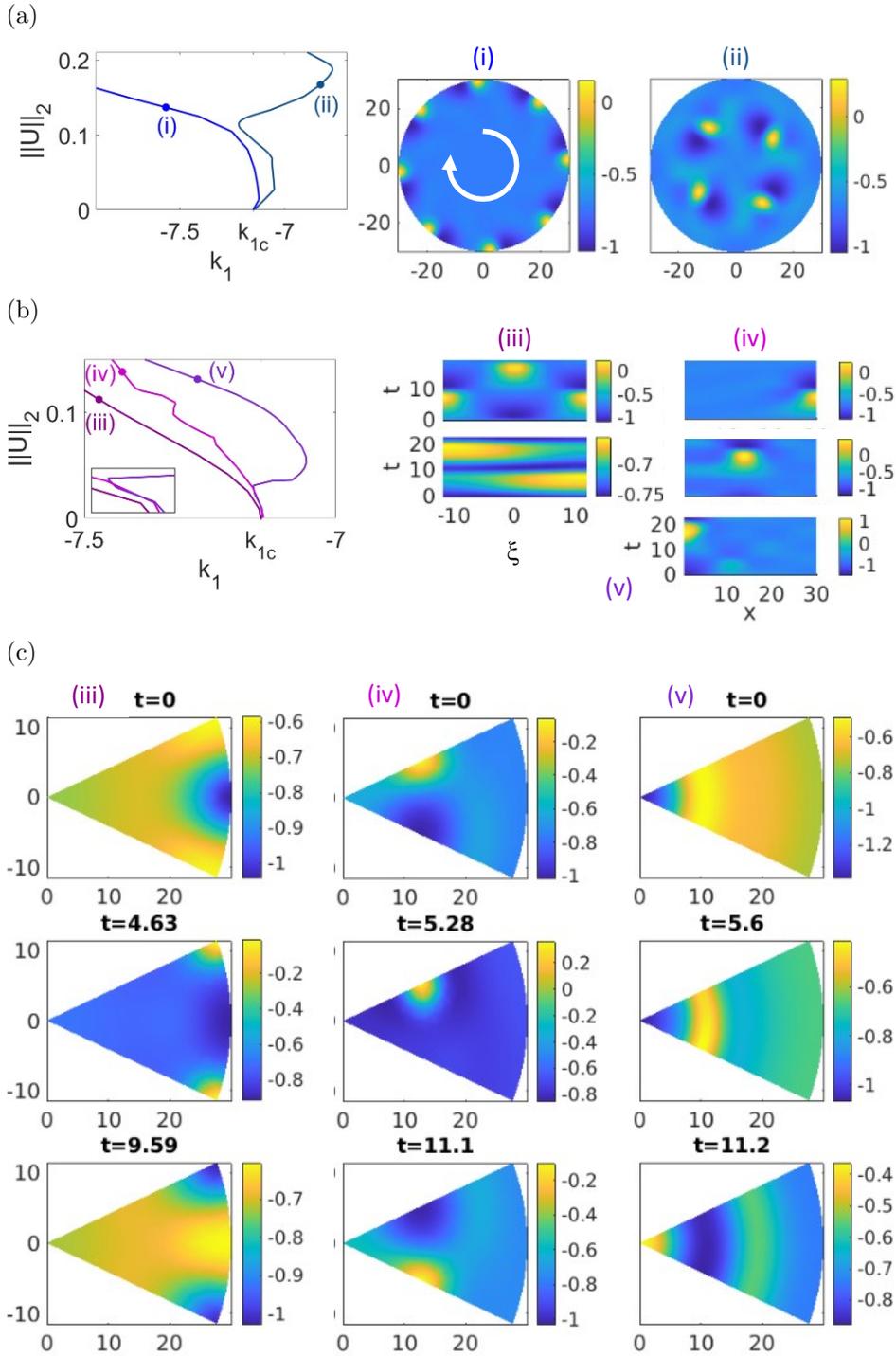

	\centering
    \btab{l l}{
        {\sm (a)}\\ \ig[width=0.8\linewidth]{"fig10a-eps-converted-to"}\\
        {\sm (b)}\\ \ig[width=0.8\linewidth]{"fig11a-eps-converted-to"}\\
        {\sm (c)}\\ \ig[width=0.8\linewidth]{"fig11b-eps-converted-to"}
        }
        \caption{(a) Bifurcation diagram on a disk with $R=30$, showing branches of rotating spot trains (RSTs) together with selected snapshots of solutions with eight wall spots (i) and four bulk spots (ii), at $k_1=-7.567$ and $k_1=-6.828$, respectively. (b) Bifurcation diagram of branches of standing oscillating spots and rings together with the corresponding space-time plots (iii)-(v) at $k_1=-7.471$, $k_1=-7.425$ and $k_1=-7.275$, respectively, showing the evolution along the perimeter $\xi$ and radius $x$ in $\pi/4$ sectors, see the snapshots in (c) taken at $t=0,\tau/4,\tau/2$, where $\tau$ is the temporal period. The axisymmetric state (v) bifurcates from the uniform state at $k_1\simeq -7.144$, see inset in (b).}	\label{fig:fig9}
\end{figure}
\begin{figure}[ht!]
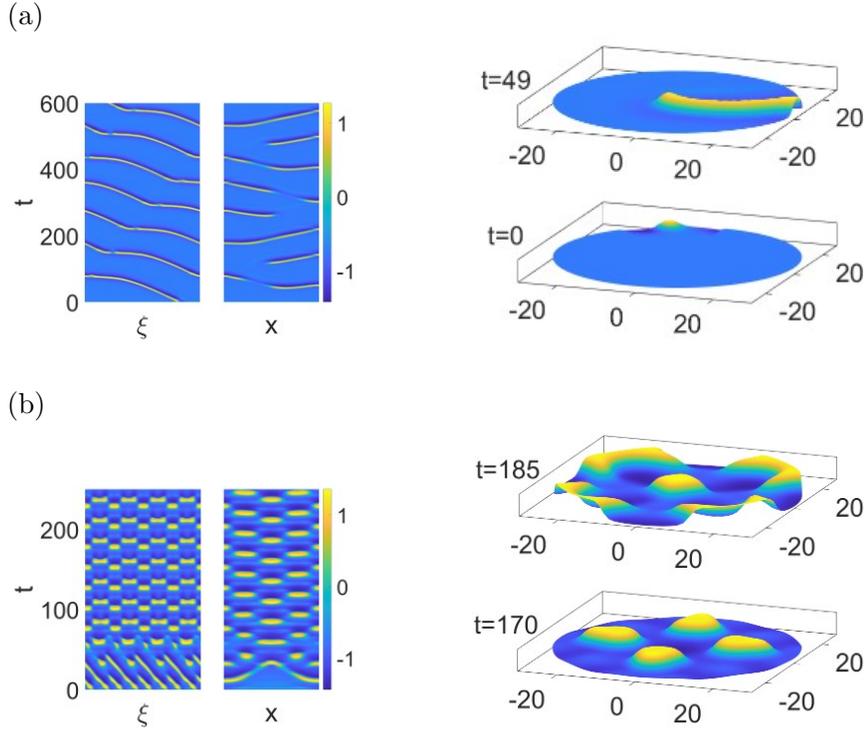

\centering
    \btab{l}{
        (a)\\ \ig[width=0.3\linewidth]{"fig12a1-eps-converted-to"}
        \hspace{0.5in}\ig[width=0.35\linewidth]{"fig12a2-eps-converted-to"}
        } 
    \btab{l}{
        \\ (b)\\ \ig[width=0.3\linewidth]{"fig12b-eps-converted-to"}
        \hspace{0.5in}\ig[width=0.35\linewidth]{"fig12c-eps-converted-to"}
        }    
    \caption{Space-time plots showing selected mixed rotating and oscillating states obtained via DNS on a disk with $R=30$ along the perimeter ($\xi$, left panels) and diameter ($x$, right panels). (a) Formation of a rotating one-arm spiral ($t=49$) from a single wall spot at $t=0$ when $k_{1}=-7.75$, and (b) formation of a $m=4$ standing oscillation at $k_{1}=-7.5$ obtained from an initial oscillating wall spot state. In both cases $k_1<k_{1c}$ and $\bU_*$ is linearly stable. See supplementary \textit{movieFigure17a} and \textit{movieFigure17b}, respectively.}
\label{fig:fig10}
\end{figure}

The greater variety of states supported on a disk is a consequence of the presence of the bulk region, whose importance is emphasized when the disk radius is increased to $R=30$. Following the same strategy, we compute branches of RSs [Fig.~\ref{fig:fig9}(a), profiles (i)-(ii)] and OSs [Fig.~\ref{fig:fig9}(b)], noting the greater variety of behavior arising from the presence of a background oscillation in the disk bulk [profiles (iii)-(v)]. One source of complexity is expected to arise from the fact that the wall and bulk oscillations may have different frequencies even if they have the same azimuthal wavenumber, resulting in the presence of a domain wall characterized by phase slips, cf.~\cite{gorman1996ratcheting,palacios,gorman2009characteristics}. 
Another arises from the tendency, with increasing forcing, of the wall states to invade the bulk. Some examples of additional states obtained via DNS are shown in Fig.~\ref{fig:fig10}. Figure~\ref{fig:fig10}(a) shows the formation of a one-arm spiral initialized by a single wall spot when the bulk is still stable, i.e., for $k_1<k_{1c}\simeq -7.149$. The spiral arm appears to be longer near the onset at $k_1=k_{1c}$ and shorter farther from it. No stable rigidly rotating spirals were observed at this radius. The second type of oscillatory behavior we have observed is an oscillation involving spots that travel outward from the disk center forming a standing pattern in the bulk that is frequency-synchronized with an $m=4$ standing oscillation at the wall [Fig.~\ref{fig:fig10}(b)]. This resembles a JO state in the radial direction associated with a standing oscillation along the wall. 
\begin{figure}[ht!]
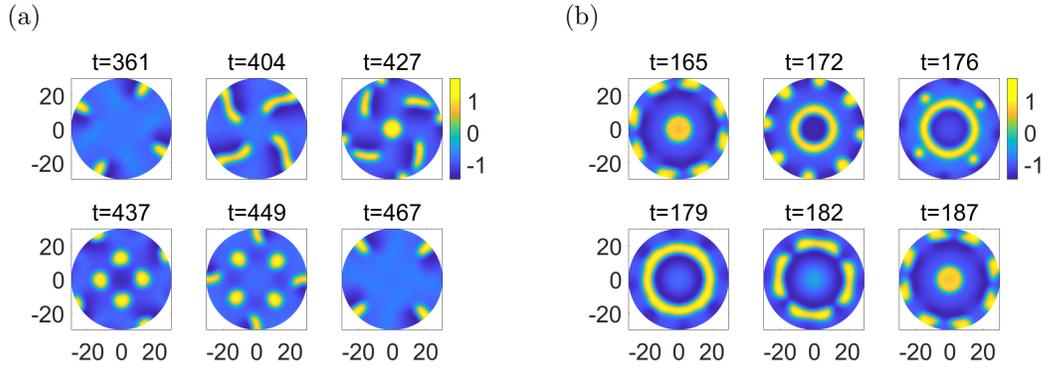

	\centering
    \btab{lll}{{\sm (a)}&\quad&{\sm (b)}\\
    \ig[width=0.42\linewidth]{"fig17-eps-converted-to"}&\quad &\ig[width=0.42\linewidth]{"fig16-eps-converted-to"}}
    \caption{Solution snapshots from DNS 
at (a) $k_1=-7.8$ (b) $k_1=-6.5$ with $R=30$. (a) The solution is dominated by a rotating wall state with wavenumber $m=4$. The wall spots invade the bulk forming a rotating 4-arm structure, which briefly excites a finite-amplitude oscillations in the disk center that generates a standing oscillation between four pairs of spots at 45$^\circ$. This oscillation dies after one oscillation period leaving again a rotating $m=4$ state at the boundary, and the process repeats. (b) The solution develops into a wall state together with an oscillation in the bulk that emits target waves traveling outwards from the center. These waves suppress the wall state when they reach the vicinity of the boundary but break up before reaching it, thereby regenerating the wall state. Throughout this process, the wall mode slowly rotates in the clockwise direction. See supplementary \textit{movieFigure18}.}
	\label{fig:fig18}
\end{figure}

Figure~\ref{fig:fig18}(a) shows the solution at the nearby parameter value 
$k_1{=}-7.8$. Here, the expansion of the wall state into the bulk excites a transient $m{=}4$ SW similar to that in Fig.~\ref{fig:fig10}(b) but one that dies away after one oscillation period before the system restores the wall state rotating along the boundary. This rotation is a consequence of the marked asymmetry of the wall state with respect to $\theta\to-\theta$ even when the four arms retract, leaving four almost radial extrusions from the boundary. 
In Fig.~\ref{fig:fig18}(b) with $k_1{=}-6.5$ and hence an unstable $\bU_*$, the solution consists of an outward traveling target wave generated by a standing oscillation in the center, which then interacts with a slowly rotating eight-spot wall state. 
\begin{figure}[tp!]
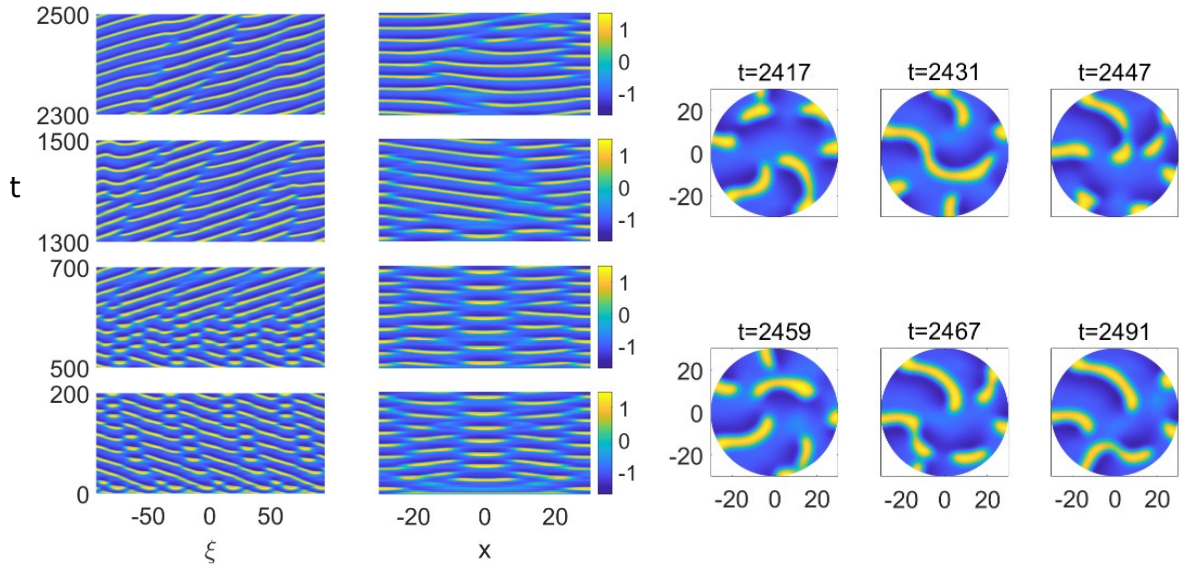

	\centering
        \ig[width=0.55\linewidth]{"fig15-eps-converted-to"} \ig[width=0.441\linewidth]{"fig15b-eps-converted-to"}
    \caption{Segments of a space-time plot (left panels) at $k_{1}=-7.1>k_{1c}\simeq -7.149$ ($\bU_*$ linearly unstable) over the time interval $t\in [0,2500]$ showing irregular changes in the direction of rotation along the boundary persisting over a long time interval. Right panels depict snapshots of the transient in the time interval $t\in [2417,2491]$, showing complex rearrangement and reconnection of spiral filaments, dominated by counterclockwise rotation along the disk boundary. See supplementary \textit{movieFigure19}.}
	\label{fig:fig10b}
\end{figure}
\begin{figure}[tp!]
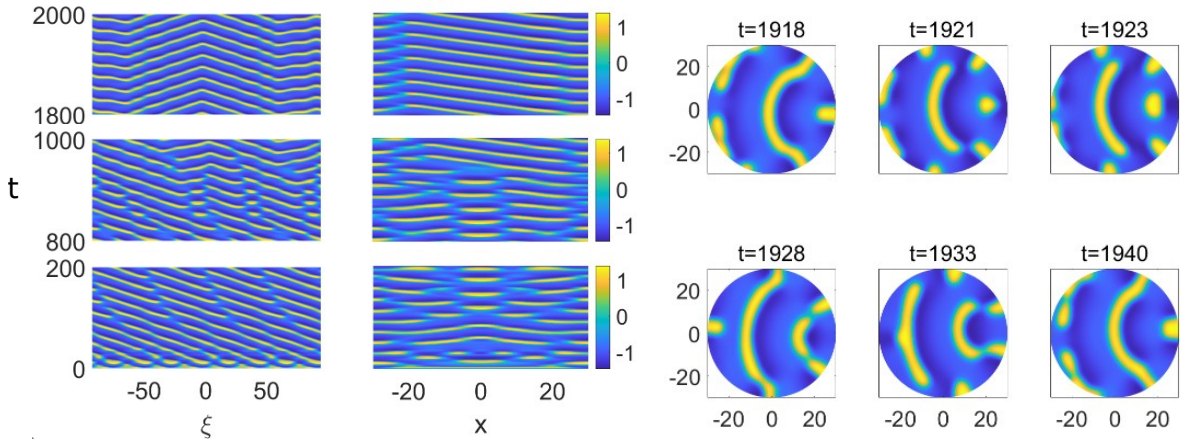

	\centering
        \ig[width=0.55\linewidth]{"fig18b-eps-converted-to"} \ig[width=0.441\linewidth]{"fig18-eps-converted-to"}
    \caption{Segments of a space-time plot (left panels) at $k_{1}=-7.0>k_{1c}$ ($\bU_*$ linearly unstable), showing that the apparently regular state shown in $t\in [0,200]$ is part of a complex transient that ultimately leads to the formation of an almost reflection-symmetric source-sink pair for counter-rotating spots along the disk boundary. Right panels depict snapshots corresponding to the time interval $t\in [1918,1940]$ showing an (almost) reflection-symmetric source-sink pair for counter-rotating wall spots. The spots collide on the right and emit a curved filament that travels to the left. This expanding filament reconnects episodically with the wall spots that travel in the opposite direction along the wall. See supplementary \textit{movieFigure20}.}
	\label{fig:fig17}
\end{figure}

In contrast, Fig.~\ref{fig:fig10b} shows the behavior at $k_{1}{=}-7.1{>}k_{1c}\simeq -7.149$. At this $k_1$ value $\bU_*$ is mildly unstable, resulting in a complex interplay between the bulk and wall modes that leads to irregular reversals in the direction of rotation of the pattern along the disk boundary. Figure \ref{fig:fig10b} also shows sample snapshots of the solution in the time interval $t{\in}[2417,2491]$ showing the complex rearrangement of spiral filaments that characterizes this transient. The solutions in Figs.~\ref{fig:fig10}(b) and \ref{fig:fig10b} 
both used a periodic rotating spot train (RST) at the perimeter as initial conditions.

We conclude this section by showing in Fig.~\ref{fig:fig17} another exotic but recurrent state at $k_1=-7$. This yields the formation, after a long transient, of a stable source-sink pair for spots traveling in opposite direction along the disk boundary. The sink serves as a source for traveling filaments in the bulk which are in turn annihilated at the source. The snapshots in the right panel of Fig.~\ref{fig:fig17} illustrate the associated dynamics over the time interval $t\in[1918,1940]$. 

\section{Discussion and conclusions}\label{sec:conclusions}
The emergence and properties of spatially localized oscillations have been a subject of considerable research since the observation of oscillons in vibrated granular media some three decades ago~\cite{umbanhowar1996localized}. 
Standard oscillons are stationary but, as shown by Yang \textit{et al.}~\cite{YZE06}, there also 
exists translating (jumping) oscillons (Fig.~\ref{fig:fig1}). These JOs were identified in DNS of a three-variable FHN model, the so-called Purwins model~\cite{schenk1997interacting}, and provide an example of not only rich and intriguing spatially localized structures but also of a distinct pattern formation mechanism~\cite{or1998spot,bode2002interaction,gurevich2004drift,gurevich2006breathing,doelman2009pulse,purwins2010dissipative,van2011planar,van2014bifurcations,teramoto2021traveling,nishiura2021matched,nishiura2022traveling}.

A key requirement for JOs is a primary finite wavenumber Hopf instability, also referred to as a wave instability, that is unavailable in two-variable reaction-diffusion models with local interaction terms where the primary Hopf bifurcation is necessarily a spatially uniform oscillation (although nonlocal coupling may yield wave instabilities in two-species reaction-diffusion equations \cite{chen2024}). The wave instability gives rise simultaneously to both traveling and standing waves~\cite{knobloch1986oscillatory} and previous work has shown, albeit in a relatively small 1D domain, that jumping oscillons emerge from a temporal modulation of the resulting traveling waves~\cite{knobloch2021origin}. Here we broadened the scope of the latter study by documenting
\bci 
\item homoclinic snaking of LSW and TPs on much larger 1D domains, as well as
\item solutions analogous to TWs, TPs and JOs on 2D disks. 
\eci 
Homoclinic snaking of oscillons and traveling pulses in 1D implies that these structures gradually grow from a single oscillon or pulse to multipulse states with more and more peaks; although these multipulse TP states turn out to be all unstable for the parameter values considered here and of significantly smaller amplitude and speed than the original stable 1TP state, they are of interest because they provide initial conditions for DNS that yield nontrivial but stable traveling structures. We expect that the multipulse traveling structures are ultimately related to the presence of a codimension-two Shil'nikov-Hopf bifurcation at nearby parameter values~\cite{hirschberg1993vsil,deng1995sil}, but in the parameter regime studied here, we do not find the large amplitude and linearly stable multipulse traveling structures identified in~\cite{yochelis2008generation,yochelis2015origin}. Instead, we find stable bound pairs of (large amplitude) TPs and JOs as in Fig.~\ref{f4}(b).

Even on 1D domains of moderate extent, we already have a large variety of states, but on 2D disks, the variety is inevitably much richer and depends strongly on the disk size. We found it useful to distinguish between dynamics along the disk perimeter and in the disk bulk; in the oscillatory regime, the latter becomes increasingly prominent as the disk size increases, while at the same time the wall-attached states behave more and more like their 1D counterparts. The bulk dynamics may be triggered even in the excitable regime via the invasion of the bulk from the boundary, that is, via a wall mode that extends sufficiently far from the boundary to excite a bulk mode. 

In 1D most of our results were obtained via continuation and bifurcation methods for (relative) equilibria and periodic orbits, supported by DNS. The details depend on the choice of boundary conditions: NBCs vs PBCs in the longitudinal direction. Most of our results were obtained with the latter, but for LSWs we used NBCs at the LSW center to reduce the computational domain. This precluded the computation of odd LSW, resulting in a single snaking branch only. Odd LSW can, of course, be computed by imposing Dirichlet boundary conditions at the LSW center, but this has not been done here. Over small disks, some primary branches (e.g., of wall states) may be found using such methods, guided by the 1D results, but here DNS methods become more important, allowing us to find states arising, for example, from the interaction between wall and bulk states. 

We believe that the approach of the present work and the resulting phenomenology will prove relevant to other three-variable systems, such as the FHN model studied in~\cite{yochelis2008generation,stich2009self} that also exhibits a finite wavenumber Hopf instability. Thus, the results obtained in this study open distinct prospects for applications to systems involving wave dynamics, ranging from intracellular waves~\cite{horning2019three,kohyama2019cell,hoffmann2025corrections,echeverria2025single,kawamura2021mathematical,sugihara2026segmented,takada2026cell,ueda2026organized}, mode-locked integrated external-cavity surface-emitting lasers~\cite{schelte2020dispersive}, through chemical reaction waves~\cite{zhabotinsky1995pattern,VE04,pena2004modulated,mikhailov2006control} to vegetation patterns~\cite{marasco2014vegetation} and predator-prey systems~\cite{chen2024}.

\appendix

\section{Numerical continuation via \pdep}\label{app:asec}
{We refer to \cite{p2pbook} and the tutorials at \cite{p2phome} for 
a description of \pdep, and specifically to \cite[Tab Applications, {\bf JOdisk}]{p2phome} containing the \pdep\ sources and software documentation for the problems studied here. In what follows, we limit ourselves to a brief review of the setup.} 

To compute solutions that are steady in a comoving frame, a.k.a. relative equilibria, e.g., traveling waves (TW) in 1D, or rotating waves (RW) on a disk, we rewrite (\ref{eq:bm0}) in the frame coming with speed $s$, i.e., we add $-s\pa_g(u,v,w)$ to the right hand side of (\ref{eq:bm0}), where 
\huga{\pa_g u=\pa_x u \text{ for TWs (1D), \quad and \quad} \pa_g u=\pa_\phi u:= (-y\pa_x+x\pa_y)u \text{ for RWs (2D).}
}
We then set the time derivatives to zero and solve the resulting nonlinear eigenvalue problem for $(U,s)$ using the phase condition
\begin{equation}\label{se1} 
    \spr{\pa_g\Uold,U}\stackrel!=0, 
\end{equation} 
where $\Uold$ is the solution from the previous  continuation step, and 
\[
\spr{U,V}\equiv\int_\varOmega \spr{U(\textbf{x}),V(\textbf{x})}\dd \textbf{x}.
\]
This minimizes the $L^2$ distance of the current step from translates of the previous step.\footnote{\label{cfoot}{ {Additionally, $k_1$ is solved for in an arclength continuation setting, where the independent parameter is the arclength along a branch; this setting allows one to follow branches that oscillate back and forth in $k_1$, and in particular to pass folds.}}} The continuation is thus orthogonal to the group orbit of translations $T_\xi U=U(x-\xi)$ (1D), resp. rotations $R_\phi U=U(R_\phi x)$ (2D), with 
\[
R_\phi=\bsm \cos\phi&\sin\phi\\-\sin\phi&\cos\phi\esm.
\]
For solutions of JO type, we retain the time derivatives and solve for both the (mean) frame speed $s$ and the oscillation period $\tau$. 
To do so, we extend the phase condition to
\begin{equation}\label{avs}
  q_H(U):=\sum_{i=1}^{m-1} \spr{\pa_g U_*, U(t_i)}\stackrel!=0, 
\end{equation} 
where $\bU_*=\bU_*(\textbf{x})$ is a reference profile (usually $\bU_H(\textbf{x})$, the spatial profile at the Hopf point) and $t_1, t_2\ldots, \tau$ are the
grid points of the time discretization. Consequently, for mTW and mRW we have the three unknowns $(U,s,\tau)$ and solve the two equations \reff{eq:bm0} and \reff{avs} together with the additional temporal phase condition 
\begin{equation}
  \int_0^\tau \spr{\ddt \Uold(t'),U(t')}\dd t'\stackrel!=0
\end{equation} 
to make the continuation orthogonal to the group orbit of time translates.  

For TWs and RWs we thus have $n_u+2$ unknowns (including $k_1$), where $n_u=3n_p$ and $n_p$ is the number of spatial discretization points, while for the mTW and mRW branches (including JOs) we have $mn_u+3$ unknowns (again including $k_1$) and $mn_u+2$ equations, where $m$ is the number of temporal discretization points. For our domains, we typically use $n_p\simeq 1000$ (1D) $n_p\simeq 3000$ (2D) discretization points in space and $m=30$ to $m=80$ in time yielding on the order of $10^5$ (1D) to $10^6$ (2D) degrees of freedom.  The predictor/corrector continuation method uses a corrector based on Newton's method and carries the danger of branch-jumping when many solutions are close together. To mitigate this, we monitor the convergence speed of the Newton loops, cf.~\cite[\S3.6]{p2pbook}. We also monitor selected eigenvalues of the linearizations to check stability and detect possible branch points, which are then localized, for subsequent branch switching. For (relative) time--periodic orbits (POs), the role of eigenvalues (for stability, bifurcation detection, and branch switching) is played by Floquet multipliers. However, robust numerical computation of Floquet multipliers is expensive, and may not converge. We therefore use (linearly implicit) DNS to check the stability of POs, in particular those in Section~\ref{sec:2D} over a disk, and to study more general dynamics, including the interaction of bulk and wall modes. 

{For the computation of localized relative equilibria $(\AL(x),\AR(x))\er^{\ri\om t}$ of the CCGL~\eqref{eq:final_ampltd} we rewrite it as a 4-component reaction-diffusion system 
\huga{\label{rcgl}
\pa_t U=D\Delta U+f(U)+s\pa_x U 
}
for the real field $U=(u_1,u_2,u_3,u_4)\equiv(\re\AL,\im\AL,\re\AR,\im\AR)$, where $s\pa_x U$ comes from imposing the spatial phase condition $q_1(U)=\spr{U,\pa_x U_0}$ with $U_0$ from the previous continuation step, and where $s$ in \reff{rcgl} stays zero (numerically $10^{-10}$) throughout. Importantly,  \eqref{eq:final_ampltd} has two independent gauge symmetries, i.e., 
if $(\AL,\AR)$ is a solution, so are $(\er^{\ri \vt_1}\AL,\AR)$ and $(\AL,\er^{\ri \vt_2}\AR)$ for any $\vt_{1,2}\in\R$. Thus we need to impose two gauge conditions, viz.,
\hugast{
\text{$\spr{(\ri \ALo,0), (\AL,0)}=0$ \quad and \quad $\spr{(0,\ri \ARo), (0,\AR)}=0$,}
}
where $(\ALo,\ARo)\in\C^2$ are the solutions from the previous step. For $U=(u_1,u_2,u_3,u_4)$ these translate into 
\def\Rho{\mathrm{P}}
\huga{
\text{$(\ri \AL,0)=\Rho_1U=(-u_2,u_1,u_3,u_4)$ \quad and \quad
$(0,\ri\AR)=\Rho_2U=(u_1,u_2,-u_4,u_3)$.} 
}
Thus we augment \reff{rcgl} with the constraints 
\huga{\label{q2}
\text{$q_2(U,\rho_1)\equiv\spr{\Rho_1 U_o,U}-\rho_1=0$ \quad and \quad $q_3(U,\rho_2)\equiv\spr{\Rho_2 U_o,U}-\rho_2=0$}.  
} 
without further modifying the rhs of \reff{rcgl}, i.e., $\rho_1, \rho_2$ are not Lagrange multipliers but just some constants. Hence we have three constraints $Q\equiv(q_1,q_2,q_3)=0$, and for continuation, we need four free parameters, for which we choose $S_g$ as the primary continuation parameter, and $\al_i, \rho_1, s$ 
as secondary parameters.} 

{For $S_g=0$, \eqref{eq:final_ampltd} has explicit pulse solutions $(\AL,\AR)=(A_{\text{LSW}},A_{\text{LSW}})$ given by \eqref{eq:clSW}, and we use these as starting points for continuation in $S_g$ of the associated pulses in \reff{rcgl}. The free $\al_i$ then yields relative equilibria in the sense of \reff{eq:clSW}, with $\Omega_{\text{LSW}}$ factored off and modifying $\al_i$ in \eqref{eq:final_ampltd}, and accordingly in \eqref{rcgl}. From these solutions of \reff{rcgl} we reconstruct the fields $(u,v,w)$ via \eqref{eq:perturb} to order $\del$ with $\del=0.035/S_g$ as approximate solutions of \reff{eq:bm0}, yielding Figs.~\ref{fig:fig3d} and~\ref{fig:fig3dphase}.} 

\section{Derivation of the amplitude equations} \label{app:amp_eq} The derivation of the amplitude equations~\eqref{eq:final_ampltd} summarized below is predicated on one key assumption - that the advection and diffusion time scales responsible for the evolution of large scale amplitude modulations are comparable. This requirement implies that the group speed $s_g$ is itself small: $s_g=\delta\,\hat{s}_g$, where $\hat{s}_g \sim \mathcal{O}(1)$.

We start by inserting~\eqref{eq:perturb} into the model system~\eqref{eq:bm0} and collecting terms in powers of the small parameter $\delta$, whose magnitude is determined by the distance from the bifurcation onset: $k_1{-}k_{1c}{=}\delta^2 \lambda$, where $\lam{=}-1$ ($\lam{=}1$) for subcritical (supercritical) branching. Since $\bU_*$ depends on $k_1$ (and thus on $\delta^2$), we also expand $u_*= u_{*c} + \delta^2 \hat u_{*} + \mathcal{O}(\delta^4)$, where $u_{*c}=u_{*}(k_{1c})$ and
\begin{align*}
    \hat u_{*}=\frac{(\Delta + k_{1c}/2)^{1/3}+(\Delta - k_{1c}/2)^{1/3}}{6 \Delta}\lambda \equiv \hat u_{*1} \lambda,\quad u_{*1} \sim \mathcal{O}\bra{1},
\end{align*}
with $(v_{*},w_{*})=(u_{*},u_{*})$. For the space-time dependent amplitudes we obtain and solve, order by order, equations of the type 
\begin{align}\label{eq:orderi}
	\mathcal{L}\bU_i=&\mathbf{R}_i,
 \end{align}
where $\mathcal{L}:={\partial_t}  \mathbb{I} - J(\bU_*)|_{k_1 = k_{1c}}$ is the linear operator related to $J$, the Jacobian of~\eqref{eq:bm0}:
\[
\mathcal{L}=\left( \begin{array}{ccc} \pa_{t} - D_u\pa^2_{x} - k_2 + 3u_{*c}^2 & k_3 & k_4 \\
	-\theta^{-1} & \pa_{t} + \theta^{-1}\bra{1-D_v\pa^2_{x}} & 0\\
-\vt^{-1} & 0 & \pa_{t} + \vt^{-1}\bra{1-D_w\pa^2_{x}}
 \end{array} \right).
\]
System~\eqref{eq:orderi} may include resonant (a.k.a. secular) terms, and their elimination is done by applying the \textit{solvability condition} (here a scalar product)
\begin{equation}\label{eq:solv}
\langle \mathcal{L}^\dagger \mathbf{N},\bU_i \rangle=0=\langle \mathbf{N},\mathbf{R}_i \rangle,
\end{equation}
where 
\[
	\langle f,g \rangle=\frac{\omega_c q_c}{\bra{2\pi}^2}\int_{0}^{2\pi/\omega_c}\int_{0}^{2\pi/q_c} \bar f g\,{\text d}x {\text d}t,
\]
and $\mathcal{L}^\dagger$ is the adjoint of $\mathcal{L}$,
\[\mathcal{L}^\dagger = \left( \begin{array}{ccc} -\pa_{t} - D_u\pa^2_{x} - k_2 +3u_{*c}^2 & -\theta^{-1} & -\vt^{-1} \\
	 k_3 & -\pa_{t} + \theta^{-1}\bra{1-D_v\pa^2_{x}} & 0\\
	 k_4 & 0 & -\pa_{t} + \vt^{-1}\bra{1-D_w\pa^2_{x}}
\end{array} \right),
\]
with null-vector
\[
\mathbf{N}=\left( \begin{array}{c} 1 \\	\dfrac{\theta k_3}{i \omega_c \theta - D_v q_c^2 -1} \\ \dfrac{\vt k_4}{i \omega_c \vt - D_w q_c^2 -1}
\end{array} \right) e^{i \omega_c t \pm i q_c x}+c.c.=\left( \begin{array}{c} 1 \\	b_2\\ b_3
\end{array} \right) e^{i \omega_c t \pm i q_c x}+c.c.
\]
We separate the contributions order by order in $\delta$ and at $\mathcal{O}(\delta)$ find the linear problem 
\begin{equation*}\label{eq:order1}
	\mathcal{L}\bU_1=\mathbf{R}_1=\left( \begin{array}{c} 0 \\ 0 \\ 0 \end{array} \right),
\end{equation*}
yielding the eigenrelations 
\begin{equation}
A_{{\rm R,L};2}{=}\frac{1}{1{+}D_v q_c^2{+}i\omega_c\theta}A_{{\rm R,L};1}
{\equiv}a_2A_{{\rm R,L};1},\quad
A_{{\rm R,L};3}{=}\frac{1}{1{+}D_w q_c^2{+}i\omega_c\vt}A_{{\rm R,L};1}{\equiv} 
a_3A_{{\rm R,L};1}.	
\end{equation}
Proceeding to $\mathcal{O} (\delta^2)$, we obtain 
\begin{align*}\label{eq:order2}
\mathcal{L}\bU_2 =& \mathbf{R}_2 = -\pa_{T_1}\bra{ \begin{array}{c} u_1 \\ v_1 \\ w_1 \end{array}}
+ 2 \mathcal{D}\pa_{X}\pa_{x} \bra{ \begin{array}{c} u_1 \\ v_1 \\ w_1 \end{array}} - 3 u_{*c} u_1^2 \bra{ \begin{array}{c} 1 \\ 0 \\ 0 \end{array}}\\
=&\bra{ \begin{array}{c} 
-\pa_{T_1} + 2i D_u q_c\pa_{X_1} \\ 
-a_2\pa_{T_1} + 2i a_2 D_v q_c \theta^{-1}\pa_{X_1}\\
-a_3\pa_{T_1} + 2i a_3 D_w q_c \vt^{-1}\pa_{X_1}\end{array}} A_{\text{L}1} e^{i(\omega_ct + q_cx)} \\&+ 
\bra{ \begin{array}{c} 
-\pa_{T_1} - 2i D_u q_c\pa_{X_1} \\ 
-a_2\pa_{T_1} - 2i a_2 D_v q_c \theta^{-1}\pa_{X_1}\\
-a_3\pa_{T_1} - 2i a_3 D_w q_c \vt^{-1}\pa_{X_1}\end{array}} A_{\text{R}1} e^{i(\omega_ct - q_cx)}
 -6u_{*c}W \bra{ \begin{array}{c} 1 \\ 0 \\ 0 \end{array}} +c.c.,
\end{align*}
where
\begin{align*}
\mathcal{D} =& \,{\rm diag} \Bra{D_u,D_v\theta^{-1},D_w \vartheta^{-1}},\\
W=&\frac{1}{2}\bra{A_{\text{L}1}^2e^{2i(\omega_ct+q_cx)}+A_{\text{R}1}^2e^{2i(\omega_ct-q_cx)}+\bar{A}_{\text{L}1}^2e^{-2i(\omega_ct+q_cx)}+\bar{A}_{\text{R}1}^2e^{-2i(\omega_ct-q_cx)}}\\
&+A_{\text{L}1}A_{\text{R}1}e^{2i\omega_ct}+\bar{A}_{\text{L}1}\bar{A}_{\text{R}1}e^{-2i\omega_ct}+A_{\text{L}1}\bar{A}_{\text{R}1}e^{2iq_cx}+\bar{A}_{\text{L}1}A_{\text{R}1}e^{-2iq_cx}+\left|A_{\text{L}1}\right|^2+\left|A_{\text{R}1}\right|^2.
\end{align*}	
Applying the solvability condition~\eqref{eq:solv} to $\mathcal{L}\bU_2=\mathbf{R}_2$ results in the hyperbolic equations
\begin{equation}\label{eq:advection}
\pa_{T_1}A_{\text{L}1} - s_g \pa_{X_1}A_{\text{L}1} = 0,\quad \pa_{T_1}A_{\text{R}1} + s_g\pa_{X_1}A_{\text{R}1} = 0,
\end{equation}
where 
\begin{align}\label{eq:group}
s_g = 2 i q_c\frac{D_u + D_v a_2 \bar{b}_2\theta^{-1} + D_w a_3 \bar{b}_3 \vt^{-1}}{1 + a_2 \bar{b}_2+a_3 \bar{b}_3} \simeq -0.074033.
\end{align}
In fact, this expression follows directly from the linear dispersion relation at $k_1=k_{1c}$, i.e., $s_g=\rm d\omega/\rm dq$ at $q=q_c$. 
Thus, spatial inhomogeneities propagate left and right at the group speed. A systematic analysis of the case $s_g={\cal O}(1)$ may be found in~\cite{KnoblochDeLucaNON3} and leads to asymptotic but {\it nonlocal} equations for the nonlinear evolution of $A_{\text{L1,R1}}$ on the time scale $T_2$.
We proceed by solving for $\bU_2$:
\begin{flalign}\label{eq:U2}
\nonumber \left( \begin{array}{c} u_2\\v_2\\w_2 \end{array}\right) =& 
\bra{ \begin{array}{c} \varphi_1\\\varphi_2\\\varphi_3 \end{array}} + \bra{\begin{array}{c} \alpha_1\\\alpha_2\\\alpha_3 \end{array}} e^{2iq_cx} 
\nonumber + \bra{\begin{array}{c} \gamma_1\\\gamma_2\\\gamma_3 \end{array}} e^{2i\omega_ct} + \bra{ \begin{array}{c} \beta_1\\\beta_2\\\beta_3 \end{array}} e^{2i(\omega_ct+q_cx)} \\
 &+ \bra{\begin{array}{c} \mu_1\\\mu_2\\\mu_3 \end{array}} e^{2i(\omega_ct-q_cx)}
\nonumber + \bra{\begin{array}{c} \zeta_1\\\zeta_2\\\zeta_3 \end{array}} e^{i(\omega_ct+q_cx)} + \bra{\begin{array}{c} \kappa_1\\\kappa_2\\\kappa_3 \end{array}} e^{i(\omega_ct-q_cx)} 
+ c.c. ,
\end{flalign}
where 
\begin{align*}
\bra{\begin{array}{c} \varphi_1\\\varphi_2\\\varphi_3 \end{array}} =& \bra{ \begin{array}{c} 1 \\ 1 \\ 1 \end{array}} \varphi \bra{\left|A_{\text{L}1}\right|^2+\left|A_{\text{R}1}\right|^2}, \quad \varphi = \frac{6 u_{*c}}{k_2 - k_3 - k_4 - 3 u_{*c}^2}, \\
\bra{ \begin{array}{c} \alpha_1\\\alpha_2\\\alpha_3 \end{array}} =& \bra{ \begin{array}{c} 1\\ a_v^{-1}\\ a_w^{-1} \end{array}} a_u  A_{\text{L}1}\bar{A}_{\text{R}1}, \quad a_u = -\frac{6 a_v a_w u_{*c}}{a_w k_3 + a_v k_4 + a_v a_w \left(4 D_u q_c^2 - k_2 + 3 u_{*c}^2 \right)}, \\
&a_v = 1+ 4D_v q_c^2,\quad a_w = 1+4D_w q_c^2, \\
\bra{\begin{array}{c} \gamma_1\\\gamma_2\\\gamma_3 \end{array}} =& \bra{ \begin{array}{c} 1 \\g_v^{-1} \\g_w^{-1} \end{array}} g_u A_{\text{L}1}A_{\text{R}1},\quad g_u = - \frac{6 g_w g_v u_{*c}}{g_w k_3 + g_v k_4 + g_v g_w \left( 3 u_{*c}^2 - k_2 + 2i \omega_c \right)}, \\
&g_v = 1+2i\theta\omega_c,\quad g_w = 1+2i\vt\omega_c, \\
\bra{\begin{array}{c} \beta_1\\\beta_2\\\beta_3 \end{array}} =& \left( \begin{array}{c} 1 \\e_v^{-1} \\e_w^{-1} \end{array}\right) e_u A_{\text{L}1}^2, \quad \bra{ \begin{array}{c} \mu_1\\\mu_2\\\mu_3 \end{array}} = \left( \begin{array}{c} 1 \\e_v^{-1} \\e_w^{-1} \end{array}\right) e_u A_{\text{R}1}^2, \\
&e_u = - \frac{3 e_w e_v u_{*c}}{e_w k_3 + e_v k_4 + e_v e_w\left( 4 D_u q_c^2 - k_2 + 3 u_{*c}^2 + 2i \omega_c \right)},\\ 
&e_v = 1+4D_v q_c^2 + 2i\theta \omega_c,\quad e_w = 1+4D_w q_c^2 + 2i\vt \omega_c,\\
\bra{\begin{array}{c} \zeta_1\\\zeta_2\\\zeta_3 \end{array}} =&  \left( \left(\begin{array}{c} 1 \\ a_2 \\ a_3 \end{array}\right) \zeta +\left(\begin{array}{c} 0 \\ a_2^2 z_v \\ a_3^2 z_w \end{array} 
\right)\right) \pa_{X_1} A_{\rm L1},\\ 
\bra{\begin{array}{c} \kappa_1\\\kappa_2\\\kappa_3 \end{array}} =& -\left( \left(\begin{array}{c} 1 \\ a_2 \\ a_3 \end{array}\right) \zeta +\left(\begin{array}{c} 0 \\ a_2^2 z_v \\ a_3^2 z_w \end{array} \right)\right) \pa_{X_1} A_{\rm R1},\\
&\zeta = \frac{z_u - a_2^2 k_3 z_v - a_3^2 k_4 z_w}{a_2 k_3 + a_3 k_4 + D_u q_c^2 - k_2 + 3 u_{*c}^2  +i\omega_c},\\
&z_u = 2 i D_u q_c - s_g,\quad z_v = 2 i D_v q_c - s_g \theta,\quad z_w = 2 i D_w q_c - s_g \vt.
\end{align*}
At $\mathcal{O} (\delta^3)$ we obtain:
\begin{align*}\label{eq:order3}
\mathcal{L}\bU_3 = \mathbf{R}_3 =& \mathcal{D}\pa^2_{X_1}\bra{ \begin{array}{c} u_1 \\ v_1 \\ w_1 \end{array} } - \bra{u_1^3 + 6 u_{*c} \hat u_{*} u_1 + 6 u_{*c}  u_1 u_2 } \bra{ \begin{array}{c} 1 \\ 0 \\ 0 \end{array}}\\
&- \pa_{T_2}\bra{ \begin{array}{c} u_1\\ v_1\\ w_1 \end{array}} + 2\mathcal{D}\pa_x\pa_{X_1}\bra{ \begin{array}{c} u_2 \\ v_2 \\ w_2 \end{array}} - \pa_{T_1}\bra{ \begin{array}{c} u_2\\ v_2\\ w_2 \end{array}}.
\end{align*}
The corresponding solvability condition is
\begin{subequations}\label{eq:amplt_slow1}
\begin{align}
\partial_{T_2} A_{\text{L1}} - s_g \pa_{X_2} A_{\text{L1}} &= \lambda \alpha A_{\text{L1}} + \beta \pa_{X_1}^2 A_{\text{L1}} + \left( \gamma |A_{\text{L1}}|^2 + \eta |A_{\text{R1}}|^2 \right) A_{\text{L1}},\\
\partial_{T_2} A_{\text{R1}} + s_g \pa_{X_2} A_{\text{R1}}&= \lambda \alpha A_{\text{R1}} + \beta \pa_{X_1}^2 A_{\text{R1}} + \left( \gamma |A_{\text{R1}}|^2 + \eta |A_{\text{L1}}|^2 \right) A_{\text{R1}}.	
\end{align}
\end{subequations}

\begin{figure}[tp!]
	\centering
	\includegraphics[width=0.75\linewidth]{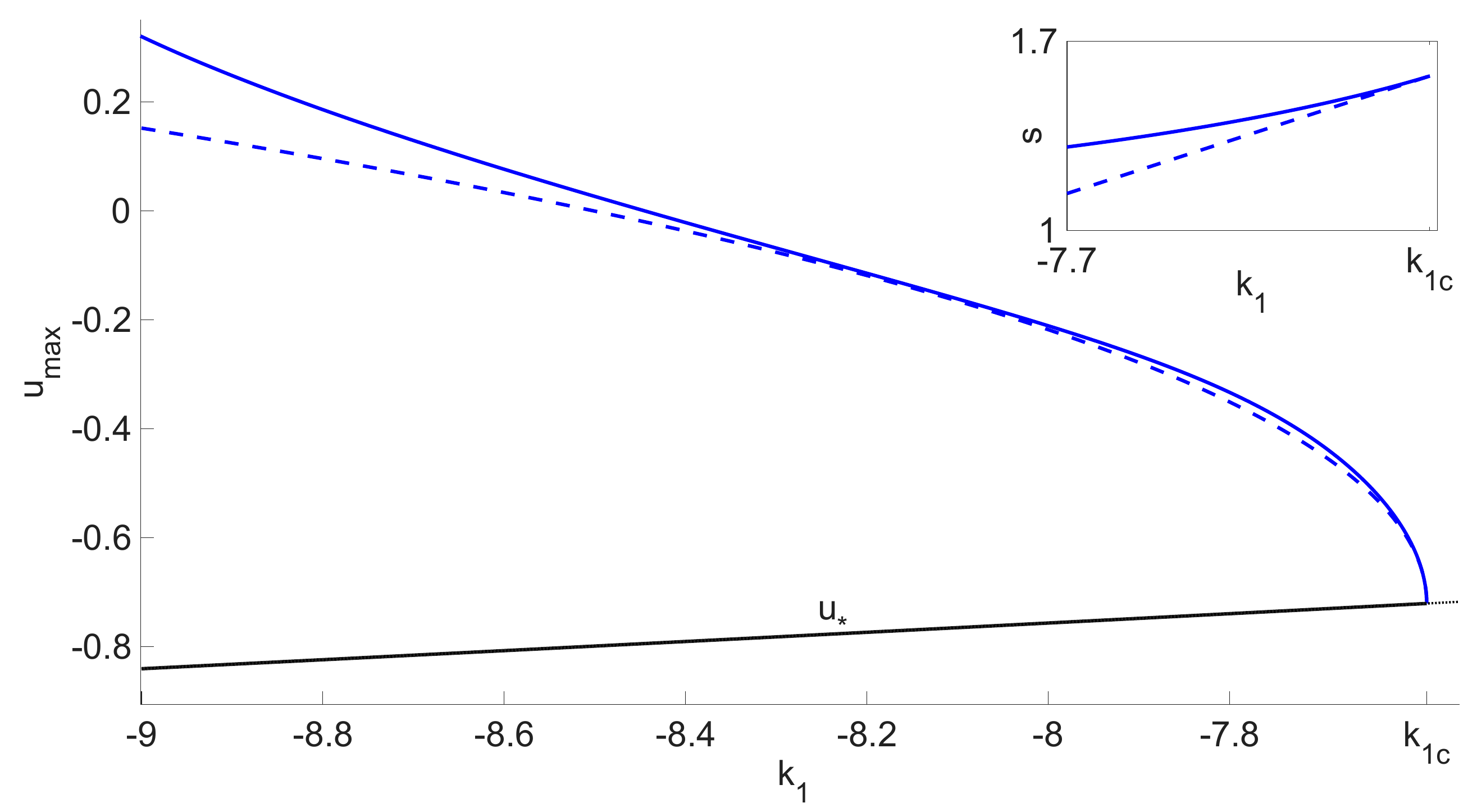}
        \caption{Bifurcation diagram of TWs obtained via continuation using AUTO (solid line) and the solution of amplitude equations (dashed line) in terms of $u_{\rm max}$ reconstructed from~\eqref {eq:TW_ampltds} and~\eqref{eq:slowamp}. The inset shows the respective speeds. A similar comparison between numerically computed SWs and LSWs and the amplitude equations is shown in Fig.~\ref{fig:fig3}.}
	\label{fig:appTW}
\end{figure}

To obtain effective amplitude equations, we employ the method of {\it reconstitution}~\cite{spiegel1981physics,roberts1985introduction}. This procedure is predicated on the existence of a description in terms of local amplitude equations (employing several solvability conditions) but is {\it nonasymptotic}. Specifically, we redefine the slow variables $X=X_1+\delta X_2$, $T=\delta^{-1}T_1+T_2$ and add a $\delta$-multiple of \eqref{eq:amplt_slow1} to \eqref{eq:advection}, obtaining 
\begin{subequations}\label{eq:amplt_slow2}
\begin{align}
\partial_{T} A_{\text{L}} - \delta^{-1}{s}_g \partial_{X} A_{\text{L}} &= \lambda \alpha A_{\text{L}} + \beta \partial_{X}^2 A_{\text{L}} + \left( \gamma |A_{\text{L}}|^2 + \eta |A_{\text{R}}|^2 \right) A_{\text{L}},\\
\partial_{T} A_{\text{R}} + \delta^{-1}{s}_g \partial_{X} A_{\text{R}}&= \lambda \alpha A_{\text{R}} + \beta \partial_{X}^2 A_{\text{R}} + \left( \gamma |A_{\text{R}}|^2 + \eta |A_{\text{L}}|^2 \right) A_{\text{R}},
\end{align}
\end{subequations}
where we have renamed $A_{\text{L1}},A_{\text{R1}}$ as $A_{\text{L}},A_{\text{R}}$, now functions of $X,T$. 
In these equations
\begin{align*}
 \quad 
\alpha=& - \frac{6 u_{*c} \hat u_{*1}}{1 + a_2 \bar{b}_2 + a_3 \bar{b}_3} \simeq 0.189 - 0.025i,\\
\beta =& \dfrac{D_u + \zeta z_u + (D_v + a_2 \bar{b}_2 (z_v (a_2 z_v+\zeta)))\theta^{-1}}{1 + a_2 \bar{b}_2 + a_3 \bar{b}_3}\\
&+\dfrac{(D_w + a_3 \bar{b}_3 (z_w (a_3 z_w+\zeta)))\vt^{-1}}{1 + a_2 \bar{b}_2 + a_3 \bar{b}_3}\simeq 4.483 - 0.724i,\\
\gamma =& -6 \dfrac{u_{*c} ( e_u +\varphi ) + 1/2}{1 + a_2 \bar{b}_2 + a_3 \bar{b}_3}\simeq 1.090 - 6.128i,\\  
\eta =& -6 \dfrac{u_{*c} \bra{a_u + g_u +\varphi } + 1}{1 + a_2 \bar{b}_2 + a_3 \bar{b}_3} \simeq 12.393 - 1.178i,
\end{align*}
$\lambda=\frac{k_1-k_{1c}}{\delta^2}=\pm 1$, and higher order terms in $\delta$ have been omitted. Observe that these equations are dominated by the advection terms when $s_g={\cal O}(1)$ and $\delta$ is small. In the present case, however, the computed value of $s_g$ is quite small and as a result equations \eqref{eq:amplt_slow2} provide a useful model for $s_g={\cal O}(\delta)$, i.e., for $\delta\sim 10^{-2}-10^{-1}$ but not too small. Of course $s_g$ has a fixed value~\eqref{eq:group} which does not vanish with approach to criticality unless a second parameter is varied at the same time. Finally, we rewrite~\eqref{eq:amplt_slow2} by 
rescaling 
$X{\rightarrow}X/\sqrt{\beta_r}, A_{\rm L, \rm R}{\rightarrow}A_{\rm L, 
\rm R} \sqrt{|\gamma_r|}$,
\begin{subequations}\label{eq:amplt_slow3}
\begin{align}
\partial_{T} A_{\text{L}} - {S}_g \partial_{X} A_{\text{L}} &= \lambda\alpha A_{\text{L}} + \bra{1 + i\tilde \beta} \partial_{X}^2 A_{\text{L}} + \Bra{ \bra{1 + i\tilde \gamma} |A_{\text{L}}|^2 + \tilde \eta |A_{\text{R}}|^2 } A_{\text{L}},\\
\partial_{T} A_{\text{R}} + {S}_g \partial_{X} A_{\text{R}}&= \lambda\alpha A_{\text{R}} + \bra{1 + i\tilde \beta} \partial_{X}^2 A_{\text{R}} + \Bra{ \bra{1 + i\tilde \gamma} |A_{\text{R}}|^2 + \tilde \eta |A_{\text{L}}|^2 } A_{\text{R}},
\end{align}
\end{subequations}
where ${S}_g = s_g/\delta\sqrt{\abs{\beta_r}}$, $\tilde \beta = \beta_i/\abs{\beta_r}$, $\tilde \gamma = \gamma_i/\abs{\gamma_r}$, $\tilde \eta = \eta/\abs{\gamma_r}$. Note that in~\eqref{eq:final_ampltd} in the main text, we have dropped the tilde symbols. Figures~\ref{fig:fig3} and \ref{fig:appTW} show the comparison between the spatially uniform and localized amplitude equations predictions (dashed lines) with numerical continuation of~\eqref{eq:bm0}. 

\section{Derivation of the linear coefficients from the dispersion relation}\label{app:disp_rel} 
The onset values $u_{*c}$, $q_c$, $\omega_c$ and the linear coefficients $s_g$, $\alpha$, $\beta$ computed in the previous Appendix~\ref{app:amp_eq}, can be derived semi-analytically from the dispersion relation for the (complex) growth rate $\sigma\equiv \sigma_r+i\omega$, mentioned at the beginning of Section~\ref{sec:1D}. This growth rate depends on $u_{*}$ (equivalently $k_1$) and the perturbation wavenumber $q^2$ and takes the form of a cubic equation: $\sigma^3 + a\sigma^2 + b \sigma + c = 0$, where $a,b,c$ are real-valued functions of $u_{*}$, $q^2$ and the original system parameters \eqref{eq:opar}. Separating the cubic equation into real and imaginary parts, we arrive at two equations:
\begin{subequations}\label{eq:disprel_reim}
\begin{align}
    &\sigma_r^3 + a \sigma_r^2 + b \sigma_r - 3 \sigma_r \omega^2 - a \omega^2  + c = 0,\\
    &3 \sigma_r^2 + 2 a \sigma_r - \omega^2 + b = 0.
\end{align}
\end{subequations}
We suppose that the onset of instability occurs at $u_{*}=u_{*c}$, $q=q_c$. These quantities are determined from the condition $\sigma_r=0$ and the requirement that $\sigma_r<0$ for all nearby $q\ne q_c$, i.e., that $\sigma_r=0$ corresponds to a maximum of $\sigma_r$. We suppose that these conditions correspond to $\omega=\omega_c$, the onset Hopf frequency.

By \eqref{eq:disprel_reim}, the conditions $\sigma_r=0$, $d\sigma_r/dq^2=0$ are equivalent to the pair of conditions
\begin{equation}\label{conds}
c-ab=\frac{d}{dq^2}(c-ab)=0
\end{equation}
or, equivalently, to the equations
\begin{subequations}
\begin{align}
    &e q^6 + f q^4 + g q^2 + h = 0,\\
    &3 e q^4 + 2 f q^2 + g = 0,
\end{align}
\end{subequations}
where $e,f,g,h$ are coefficients that depend on the system parameters. These conditions determine the onset wavenumber,
\begin{equation}
    q_c^2 = \frac{1}{2}\frac{9 e h - f g}{f^2 - 3 e g},
\end{equation}
together with a lengthy expression for $u_{*c}$ which we omit. These results also yield the onset Hopf frequency in the form $\omega_c^2=b_c=c_c/a_c$. We have verified these expressions using the numerically determined values of $u_{*c}$ and $q_c$.

Next, we expand all dispersion relation coefficients in powers of $q^2 - q_c^2$:
\begin{equation}
    \left(\begin{array}{c}  a \\ b \\ c \\ \sigma_r \\ \omega \end{array}\right) = 
    \left(\begin{array}{c}  a_c \\ b_c \\ c_c \\ \sigma_c \\ \omega_c \end{array}\right) + 
    \left(\begin{array}{c}  a_{q^2} \\ b_{q^2} \\ c_{q^2} \\ \sigma_{q^2} \\ \omega_{q^2} \end{array}\right) (q^2 - q_c^2) +
    \frac{1}{2}\left(\begin{array}{c}  a_{q^4} \\ b_{q^4} \\ c_{q^4} \\ \sigma_{q^4} \\ \omega_{q^4} \end{array}\right) (q^2 - q_c^2)^2 + h.o.t.,
\end{equation}
where $\sigma_c = 0$, and $a,b,c,\omega_c$ are all known. We transform this expansion into an expansion in $q - q_c$:
\begin{equation}
    \left(\begin{array}{c}  a \\ b \\ c \\ \sigma_r \\ \omega \end{array}\right) = 
    \left(\begin{array}{c}  a_0 \\ b_0 \\ c_0 \\ \sigma_0 \\ \omega_0 \end{array}\right) + 
    \left(\begin{array}{c}  a_1 \\ b_1 \\ c_1 \\ \sigma_1 \\ \omega_1 \end{array}\right) (q - q_c) +
    \left(\begin{array}{c}  a_2 \\ b_2 \\ c_2 \\ \sigma_2 \\ \omega_2 \end{array}\right) (q - q_c)^2 + h.o.t.,
\end{equation}
where  
$$
\mu_0 = \mu_c,\quad \mu_1 = 2 q_c \mu_{q^2},\quad \mu_2 = \mu_{q^2} + 2 q_c^2 \mu_{q^4}
$$
with $\mu \equiv a,b,c,\sigma,\omega$.
We insert these expressions into \eqref{eq:disprel_reim} and separate into powers of $q - q_c$. At zeroth order we recover the onset conditions. At order $q - q_c$ we obtain expressions for $\left(\sigma_1,\omega_1\right)$ and find that $({\rm d} \sigma_r/{\rm d} q,{\rm d} \omega/{\rm d}q)|_{k_{1c},q_c}=(0,s_g)$, as discussed in Section~\ref{sec:1D}. Numerically, $\left(\sigma_1,\omega_1\right) \simeq (0,-0.074)$. Continuing to second order in $q-q_c$, we find that $\left(\sigma_2,\omega_2\right) \simeq -(4.483,-0.724)$ which agrees with the values $\left(\beta_r,\beta_i\right)$ found in Appendix~\ref{app:amp_eq} using multiple scale analysis as well as with the numerical values of $-\frac{1}{2}({\rm d}^2 \sigma_r/{\rm d} q^2,{\rm d}^2\omega/{\rm d}q^2)|_{k_{1c},q_c}$ corresponding to Fig.~\ref{fig:fig2}(b).

To find the coefficient $\alpha$, we follow a similar process, but expand the coefficients and the real and imaginary parts of the dispersion relation in the control parameter $k_1$. The resulting values of $(\alpha_r,\alpha_i)$ now correspond to $\left(\sigma_{1},\omega_{1}\right) \equiv ({\rm d} \sigma_r/{\rm d} k_1,{\rm d}\omega/{\rm d}k_1)|_{k_{1c},q_c}$.

\bibliographystyle{siamplain}

\end{document}

%% file: shared.tex
\newcommand{\btab}[2]{\begin{tabular}{#1}#2\end{tabular}}
\def\sm{\small}
\newlength{\tew}

\def\ig{\includegraphics}
\def\Re{{\rm Re}}\def\Im{{\rm Im}}
\def\medskip{}\def\bigskip{}

\usepackage{tikz}
 \def\Uold{U_{{\rm old}}}

\newcommand{\bU}{\mathbf{U}}

\newcommand{\bra}[1]{\left(#1\right)} \newcommand{\Bra}[1]{\left[#1\right]}

\usepackage{multirow}\usepackage{pstricks}\usepackage{fancyvrb}
\usepackage{listings,paralist, longtable}
\def\im{{\rm Im}}
\usepackage{hyperref}
\usepackage{setspace}

\def\re{{\rm Re}}\def\im{{\rm Im}}

\def\AL{A_{\text{L}}}\def\AR{A_{\text{R}}}
\def\ALo{A_{\text{L}0}}\def\ARo{A_{\text{R}0}}

\usepackage{lipsum}
\usepackage[mathlines]{lineno}
\usepackage{amsmath}
\usepackage{amsfonts}
\usepackage{graphicx}
\usepackage{epstopdf}
\usepackage{algorithmic}
\usepackage{ulem}
\usepackage{tabularray}
\usepackage{makecell}
\usepackage{enumitem}
\ifpdf
  \DeclareGraphicsExtensions{.eps,.pdf,.png,.jpg}
\else
  \DeclareGraphicsExtensions{.eps}
\fi

\newsiamremark{remark}{Remark}
\newsiamremark{hypothesis}{Hypothesis}
\crefname{hypothesis}{Hypothesis}{Hypotheses}
\newsiamthm{claim}{Claim}

\headers{Localized spatiotemporal RD patterns}{Knobloch, Modai, Uecker, and Yochelis}

\title{Localized spatiotemporal reaction-diffusion patterns on a line and a disk arising from a subcritical finite wavenumber Hopf instability
\thanks{Submitted to the editors on \today.
\funding{This work was supported by grant no. 2022072 from the United States--Israel Binational Science Foundation (BSF), Jerusalem, Israel.}}}

\author{Edgar Knobloch\thanks{Department of Physics, University of California at Berkeley, Berkeley, CA 94720, USA.} 
\and Saar O. Modai\thanks{Department of Physics, Ben-Gurion University of the Negev, Be'er Sheva 8410501, Israel.}
\and Hannes Uecker\thanks{Institut f\"ur Mathematik, Universit\"at Oldenburg, D26111 Oldenburg, Germany.}
\and Arik Yochelis\thanks{Swiss Institute for Dryland Environmental and Energy Research, Blaustein Institutes for Desert Research, Ben-Gurion University of the Negev, Sede Boqer Campus, Midreshet Ben-Gurion 8499000, Israel, and Department of Physics, Ben-Gurion University of the Negev, Be'er Sheva 8410501, Israel (\email{yochelis@bgu.ac.il})}.}

\usepackage{amsopn}


%% file: hudef.tex
\def\bexe{\begin{exercise}}\def\eexe{\eex\end{exercise}}
\def\bsol{\begin{solution}}\def\esol{\eex\end{solution}}
\def\bexa{\begin{example}}\def\eexa{\end{example}}
\def\brem{\begin{remark}}\def\erem{\end{remark}}
\def\bthm{\begin{theorem}}\def\ethm{\end{theorem}}
\def\blem{\begin{lemma}}\def\elem{\end{lemma}}
\def\bcor{\begin{corollary}}\def\ecor{\end{corollary}}
\def\bdefi{\begin{definition}}\def\edefi{\end{definition}}
\newcommand{\IDEA}{\textbf{Idea of the Proof.} }

\newcommand{\abs}[1]{\lvert#1\rvert}

\def\bmip{\begin{minipage}{\textwidth}}\def\emip{\end{minipage}}
\def\huga#1{\begin{gather} #1 \end{gather}}
\def\hugast#1{\begin{gather*} #1 \end{gather*}}

\newcommand{\R}{{\mathbb R}}
\newcommand{\C}{{\mathbb C}}

\def\e{{\rm e}}

\def\om{\omega}

\def\vt{\vartheta}
\def\pa{{\partial}}\def\lam{\lambda}
\newcommand{\bi}{\begin{itemize}}\newcommand{\ei}{\end{itemize}}
\newcommand{\ben}{\begin{enumerate}}\newcommand{\een}{\end{enumerate}}
\newcommand{\bce}{\begin{center}}\newcommand{\ece}{\end{center}}
\newcommand{\bci}{\begin{compactitem}}\newcommand{\eci}{\end{compactitem}}
\newcommand{\bcen}{\begin{compactenum}}\newcommand{\ecen}{\end{compactenum}}
\newcommand{\bcena}{\begin{compactenum}[(a)]}
\newcommand{\reff}[1]{(\ref{#1})}

\newcommand{\spr}[1]{\left\langle #1 \right\rangle}

\newcommand{\barr}{\begin{array}}\newcommand{\earr}{\end{array}}
\newcommand{\bpm}{\begin{pmatrix}}\newcommand{\epm}{\end{pmatrix}}
\newcommand{\bsm}{\left(\begin{smallmatrix}}
\newcommand{\esm}{\end{smallmatrix}\right)}
\newcommand{\ba}{\begin{array}}\newcommand{\ea}{\end{array}}
\def\dd{\, {\rm d}}\def\ri{{\rm i}}

\def\er{{\rm e}}
\def\re{{\rm Re\,}}\def\im{{\rm Im\,}}
\def\om{\omega}
\def\ddt{\frac{\rm d}{{\rm d}t}}
\def\del{\delta}

\def\eex{\hfill\mbox{$\rfloor$}}

\def\Re{{\rm Re}}\def\Im{{\rm Im}}

\def\al{\alpha}

\def\bd{\begin{displaymath}} \def\ed{\end{displaymath}}
\def\ba{\begin{array}} \def\ea{\end{array}}


%% file: p2pdef.tex
\def\pdep{{\tt pde2path}}